\documentclass[preprints,article,accept,moreauthors,pdftex,10pt,a4paper]{mdpi} 

\firstpage{1} 
\pubvolume{xx}
\issuenum{1}
\articlenumber{5}
\pubyear{2018}
\copyrightyear{2018}
\history{Received: date; Accepted: date; Published: date}

\makeatletter
\g@addto@macro{\UrlBreaks}{%
\do\/\do\a\do\b\do\c\do\d\do\e\do\f%
\do\g\do\h\do\i\do\j\do\k\do\l\do\m%
\do\n\do\o\do\p\do\q\do\r\do\s\do\t%
\do\u\do\v\do\w\do\x\do\y\do\z%
\do\A\do\B\do\C\do\D\do\E\do\F\do\G%
\do\H\do\I\do\J\do\K\do\L\do\M\do\N%
\do\O\do\P\do\Q\do\R\do\S\do\T\do\U%
\do\V\do\W\do\X\do\Y\do\Z}
\makeatother

\Title{The Wavelength-shifting Optical Module}

\usepackage[english]{babel}
\usepackage{units}
\newcommand{\subtitle}[1]{%
  \posttitle{%
    \par\end{center}
    \begin{center}\Large#1\end{center}
    \vskip0.5em}%
}
\usepackage[separate-uncertainty=true]{siunitx}

\usepackage[final]{changes}
\definechangesauthor[name="Anna Pollmann", color=red]{AP}
\definechangesauthor[name="John Rack-Helleis", color=magenta]{JRH}
\definechangesauthor[name="Sebastian Boeser", color=violet]{SB}

\Author{Benjamin Bastian-Querner$^{1}$ \orcidK{},
Lucas S. Binn$^{2}$, 
Sebastian B\"oser$^{2}$ \orcidB{}, 
Jannes Brostean-Kaiser$^{1,3}$, 
Dustin Hebecker$^{1,3}$ \orcidE{}, 
Klaus Helbing$^{4}$ \orcidC{}, 
Timo Karg$^{3}$ \orcidD{}, 
Lutz K\"opke$^{2}$, 
Marek Kowalski$^{1,3}$ \orcidF, 
Peter Peiffer$^{2}$, 
Anna Pollmann$^{4}$* \orcidA{},   
John Rack-Helleis$^{2}$ \orcidH{}, 
Martin Rongen$^{2}$ \orcidG{}, 
Lea Schlickmann$^{2}$ \orcidI{}, 
Florian Thomas$^{2}$ \orcidJ{},  
Anna Vocke$^{2}$ 
}

\AuthorNames{Benjamin Bastian-Querner, Lucas S. Binn, Sebastian B\"oser, Jannes Brostean-Kaiser, Dustin Hebecker, Klaus Helbing, Timo Karg, Lutz K\"opke, Marek Kowalski, Peter Peiffer, Anna Pollmann, Yuriy Popovych, John Rack-Helleis, Martin Rongen, Lea Schlickmann,  Nick Jannis Schmeisser, Florian Thomas, Anna Vocke}

\address{\\
$^{1}$ \quad Institut f\"ur Physik, Humboldt-Universit\"at zu Berlin, 12489 Berlin, Germany\\
$^{2}$ \quad Institute of Physics, University of Mainz, Staudinger Weg 7, 55099 Mainz, Germany\\
$^{3}$ \quad Deutsches Elektronen-Synchrotron DESY, 15738 Zeuthen, Germany\\
$^{4}$ \quad Department of Physics, University of Wuppertal, 42119 Wuppertal, Germany\\
}

\corres{Correspondence: anna.pollmann@uni-wuppertal.de, wom@desy.de}

\addto\extrasenglish{%

}

\abstract{%
The Wavelength-shifting Optical Module (WOM) is a novel photosensor concept for the instrumentation of large detector volumes with single-photon sensitivity. The key objective is to improve the signal-to-noise ratio which is achieved by decoupling the photosensitive area of a sensor from the cathode area of its photomultiplier tube (PMT). 
The WOM consists of a transparent tube with two PMTs attached to its ends. The tube is coated with wavelength-shifting paint absorbing ultra-violet photons with nearly 100\% efficiency. 
Depending on the environment, e.g. air (ice), up to 73\% (41\%) of the subsequently emitted optical photons can be captured by total internal reflection and propagate towards the PMTs where they are recorded.
The optical properties of the paint, the geometry of the tube and the coupling of the tube to the PMTs have been optimized for maximal sensitivity based on theoretical derivations and experimental evaluations. Prototypes were built to demonstrate the technique and to develop a reproducible construction process. 
Important measurable characteristics of the WOM are the wavelength dependent effective area, the transit time spread of detected photons and the signal-to-noise ratio. The WOM outperforms bare PMTs especially with respect to the low signal-to-noise ratio with an increase of a factor up to 8.9 in air (5.2 in ice). Since the gain in sensitivity is mostly in the UV-regime, the WOM is an ideal sensor for Cherenkov and scintillation detectors.
}


\begin{document}

\section{Introduction \label{sec:intro}}

Many detectors for rare events in particle and astroparticle physics require large interaction volumes, ranging up to cubic kilometers, in order to achieve a reasonable detection rate. However, the larger the volume, the smaller the feasible density of instrumentation. 
A viable method to read out detectors with transparent target media is the detection of optical photons emitted by incident particles (or their secondaries) in form of Cherenkov emission or scintillation light.
The prevailing sensor type, which is capable of detecting \emph{single photons}, are photomultiplier tubes (PMT). They are, for example, deployed in neutrino telescopes, neutrino detectors and dark matter experiments such as IceCube \cite{icecube}, Super-Kamiokande \cite{FUKUDA2003418}, Borexino \cite{ALIMONTI2009568}, SNO \cite{SNO:1999crp}, JUNO \cite{Juno}, XENON\cite{APRILE2012573}, LUX \cite{AKERIB2013111}, DARWIN \cite{Aalbers_2016}, and many others. 
The photosensitive area of the PMT scales approximately linearly with the area of its photocathode, so does the dark current -- the dominant source of noise. 
The peak quantum efficiency of conventional PMTs is around $450\,\mathrm{nm}$ which is suboptimal for Cherenkov light as well as for scintillation emissions from common liquid scintillators based on linear alkylbenzene (LAB) at approximately $\unit[340]{nm}$ \cite{LAB}, or liquid noble gases such as  Xenon or  Argon at around $\unit[178]{nm}$ and $\unit[128]{nm}$, respectively.

\begin{figure}
\centering
\includegraphics[width=0.9\textwidth]{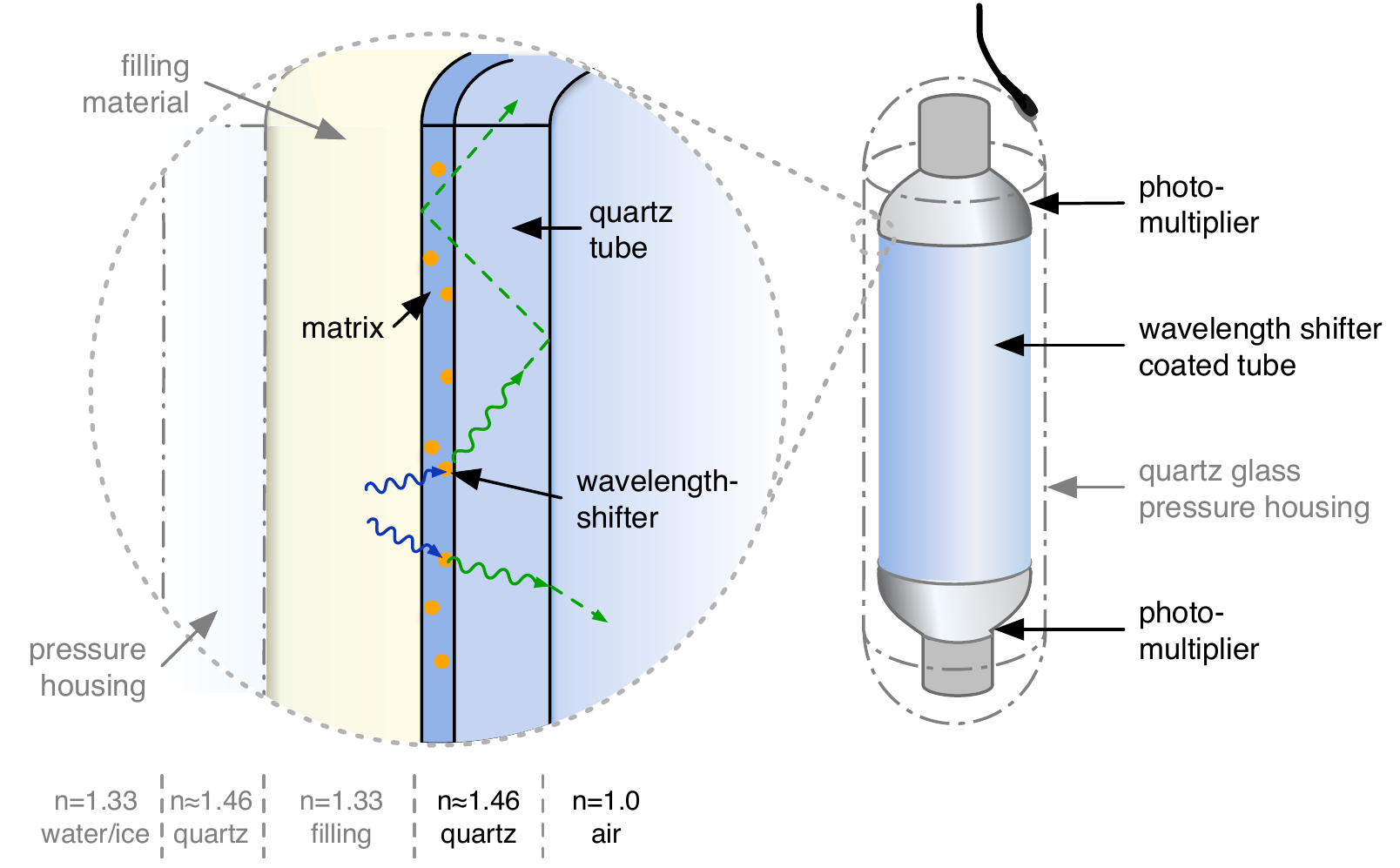}
\caption{\label{WOM-schematic}A schematic drawing of the WOM. UV photons are absorbed in the WLS paint layer and are re-emitted as optical photons. If the emission angle of the photons is larger than the critical angle, they are trapped by total internal reflection and are guided along the tube to small, low-noise PMTs.}
\end{figure}

The Wavelength-shifting Optical Module (WOM) has been developed as an alternative sensor for large volume detectors. The WOM, schematically shown in \autoref{WOM-schematic}, consists of a transparent tube with photosensors attached to its ends. %
The tube is coated with a paint containing wavelength-shifting (WLS) organic  luminophores and acts as a light collector. The delayed isotropic emission of the WLS molecules circumvents 
the conservation of phase space as required by Liouville's theorem (etendue), that otherwise forbids 
the concentration of a homogeneous light distribution on a smaller area using refractive and reflective optics.

The performance of the WOM is determined by a number of efficiency factors which are discussed throughout this work. The overall efficiency of the WOM $\epsilon_{\mathrm{tot}}$ is defined as the number of detected photons $N_{\rm det}$ compared to the number of incident photons $N_{\rm inc}$ as follows:
\begin{equation}
    \begin{aligned}
        \epsilon_{\mathrm{tot}}  = \frac{N_{\rm det}}{N_{\rm inc}} = 
        \epsilon^{\mathrm{WLS}}_{\mathrm{LY}}\left(\lambda_{\rm inc}\right)
         \cdot \epsilon_{\mathrm{TIR}} \left(x_0\right) 
         \cdot \epsilon_{\mathrm{TP}} \left(z\right) 
         \cdot \epsilon^{\mathrm{PMT}}_{\mathrm{QE}} \left(\lambda_{\rm em}\right)
         \cdot \epsilon_{\mathrm{IF}}  
         \cdot \epsilon_{\mathrm{TM}}\left(\lambda_{\rm inc}\right)
        \label{eqn:master_formula}
    \end{aligned}
\end{equation}

UV photons incident on the WLS tube are absorbed in the paint layer and re-emitted isotropically as optical photons with the light yield  $\epsilon^{\mathrm{WLS}}_{\mathrm{LY}}$.
This efficiency depends on the wavelength of the incident photons $\lambda_{\rm inc}$ as well as the thickness of the paint layer $d$ and the concentration of wavelength-shifting molecules $c_{\rm WLS}$. Values close to unity can be achieved over a large part of the absorption spectrum. 
A fraction $\epsilon_{\mathrm{TIR}}$ of the re-emitted photons has an angle with the tube wall smaller than the critical angle for total internal reflection (TIR).
The radial symmetry of the configuration ensures that the photon incidence angle on the surface remains constant for each interaction. Therefore subsequent surface interactions also fulfill the TIR criterion, such that the photons are trapped and guided towards the ends of the tube, see \autoref{fig:ray_tracing}.
The TIR fraction $\epsilon_{\mathrm{TIR}}$ depends on the fraction of the tubes radius $x_0$, at which the photons are emitted, as well as on the 
refractive indices of the tube $n_{\rm tube}$ and the environment $n_{\rm env}$ which define the critical angle.

While the captured photons are propagating towards the ends of the tube, they can be absorbed or scattered out of the material.
This attenuation effect, denoted as $\epsilon_{\mathrm{TP}}$, depends on the distance $z$ between the emission point and the end of the tube as well as the scattering $\lambda_{\rm scatt}$ and absorption length $\lambda_{\rm abs}$ (merged into the attenuation length $\lambda_{\rm att}$).
Attached to the tube are PMTs which are characterized by their wavelength dependent quantum efficiency $\epsilon_{\mathrm{QE}}^{\mathrm{PMT}}$ which depends on the wavelength of the re-emitted photons $\lambda_{\mathrm{em}}$. 
A small fraction of photons is lost at the tube-PMT interface denoted as $\epsilon_{\mathrm{IF}}$.

In some environments it is necessary to encapsulate the WOM with a  housing{, e.g. in deep water or ice to protect the WOM from pressure}. The probability for transmission into the housing through an optional filling into the WOM $\epsilon_{\mathrm{TM}}$ depends 
on the wavelength $\lambda_{\rm inc}$ of the incident photons and the refractive indices $n_{\rm env},n_{\rm housing},n_{\rm fill}$ of the materials enclosing the WLS tube. 

All these efficiencies, described above, are used to characterize the WOM in this paper. 
As a baseline in this paper, the following \emph{prototype design} is discussed: the tube, coated on the inside surface with wavelength-shifting paint, is \SI{700}{\mm} long, made of quartz \cite{hsq300} with an outer diameter of \SI{60}{\mm} and a wall thickness of \SI{2.5}{\mm}. Two 3.5 inch PMTs \cite{hamamatsu} are attached to the ends with optical gel \cite{eljen}. An optional quartz pressure vessel\footnote{{The dimension of this pressure vessel are given by availability in the laboratory. The dimensions of the WOM tube and the PMT diameters are derived from these.}} is added around the WOM with \SI{1200}{\mm} length, \SI{110}{\mm} outer diameter and \SI{10}{\mm} wall thickness. Dimensions and materials of any setup discussed in this text corresponds to this prototype design -- apart from a few exceptions which are described accordingly. 

In this paper, the different optical processes required to describe and optimize the WOM efficiency are discussed in \autoref{sec:perfomance}. 
The optimized production of the main component of the WOM, the coating of the wavelength-shifting tube, is described in \autoref{sec:paint}.
Finally, characterization and performance measurements of the WOM are presented in \autoref{sec:efficiency_measurement}.


\section{Performance factors\label{sec:perfomance}}

The efficiencies of the WOM components, which make up the overall efficiency of the WOM, introduced in \autoref{eqn:master_formula}, are derived in this chapter.


\subsection{Total internal reflection\label{sec:TIR}}


\begin{figure}
\centering
\includegraphics[width=0.8\textwidth]{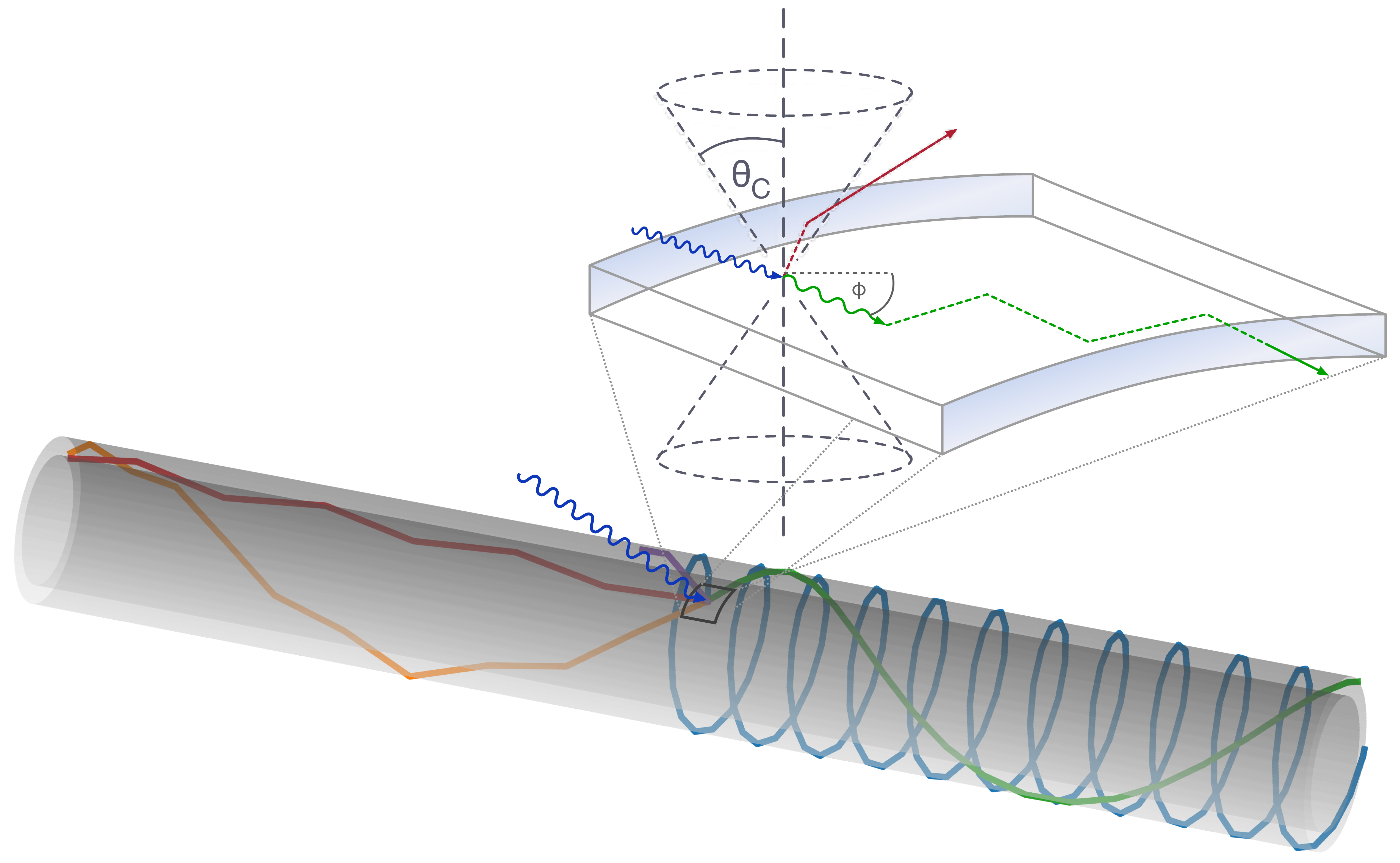}
\caption{{Visualization} of  several photons paths \deleted[id=AP]{(green)} {propagating in the wall of the WOM tube simulated with} a GPU-based ray tracing algorithm \cite{Thomas:2019}. The photons are generated inside the paint layer of the tube {in response to \replaced[id=AP]{the}{an} incident UV photon\added[id=AP]{(s)} (blue)}. Their attenuation probability depends on the effective path length which in turn depends on the emission angle.
With the inset a UV photon (blue) is shown, which is absorbed and re-emitted in the material (green) or lost (red). The critical angle for the re-emitted photon defines a \emph{loss cone} (grey {dashed lines}).}
\label{fig:ray_tracing}
\label{fig:light_trapping}
\end{figure}

The concept of the WOM evolves around the idea of total internal reflection of photons if the difference of refractive indices between the tube $n_{\rm tube}$ and the surrounding environment $n_{\rm env}$ is large\footnote{Since a hollow tube is coated on one surface side, it is the lower of the refractive indices of the wavelength-shifting paint $n_{\rm paint}$ and the coated substrate $n_{\rm substrate}$ that determines the critical angle, i.e. $n_{\rm tube} = {\rm min}(n_{\rm paint},n_{\rm substrate})$}. Using the Fresnel equation for the reflectivity of an electromagnetic wave $R(\theta, n_{\rm paint},n_{\rm env})$ for a re-emitted photon incident on the surface, the fraction of captured photons after $N$ interactions with the surface can be determined as

\begin{equation}
  \epsilon_{\rm TIR} = \frac{1}{2}\int \left( R(\theta, n_{\rm paint},n_{\rm env})\right)^N \sin\theta\,d\theta
  \label{eqn:capture}
\end{equation}

where $\theta$ is the angle of the photon trajectory $\vec{p}$ to the surface normal $\vec{n}$, thus $\cos\theta = \vec{p}\cdot\vec{n}$. In contrast to mirroring surfaces that achieve typical average reflectance of 90-99\% if employed over larger wavelength range, the TIR above the critical angle $\sin\theta_{\rm crit} = {n_{\rm env}}/{n_{\rm tube}}$ reaches 100\% reflectance independent of the wavelength. In the limit of a large number of interactions, the reflectance $R$ can be approximated by a step-function $\Theta$ and \autoref{eqn:capture} simplifies to 

\begin{equation}
  \lim\limits_{N\to\infty} \varepsilon_{\rm TIR}  = \frac{1}{2}\int \Theta(\cos\theta - \cos\theta_{\rm crit}) \sin\theta\,d\theta  = \cos \theta_{\rm crit} \label{eqn:TIR}
\end{equation}

The critical angle thus defines two sharp loss cones while all other photons are captured, see \autoref{fig:light_trapping}. In the configurations discussed here, neither the tube material nor the wavelength-shifting paint or surrounding environment exhibit strong dispersion, so that $\epsilon_{\rm TIR}  \simeq \unit[73]{\%}$ ($\theta_{\rm crit} \simeq 43.2^\circ$) for a WOM tube (${n_\mathrm{tube}} \simeq 1.46$) operated in air ($n \simeq 1.0$) and $\epsilon_{\rm TIR}  \simeq \unit[41]{\%}$ ($\theta_{\rm crit} \simeq 65.6^\circ$) for a WOM tube operated in water ($n \simeq 1.33$).
The surface curvature given by the radius of the tube $r_{\rm tube}$ has been neglected which is a good approximation as long as the radius of the emission point $r_{\rm em}$ is close to the surface $r_{\rm tube} - r_{\rm em} \ll r_{\rm tube}$ (denoted as \emph{no-curvature} approximation later on). 
In resorting to the ray-like description of reflectance, evanescence was also neglected which is a good approximation as long as the emission point is further away from the surface than the typical wavelength $r_{\rm tube} - r_{\rm em} \gg \lambda_{\rm em}$. 
Both conditions are well satisfied in the given configuration with typical emission wavelength of \SIrange{400}{500}{\nm}, coating thicknesses $r_{\rm tube} - r_{\rm em}$ in the range of several micrometers and tube radii $r_{\rm tube}$ of several centimeters. The TIR efficiency $\epsilon_{\rm TIR}$ is thus independent of the tube radius $r_{\rm tube}$ in good approximation. 

\begin{figure}
\centering
\includegraphics[width=\textwidth]{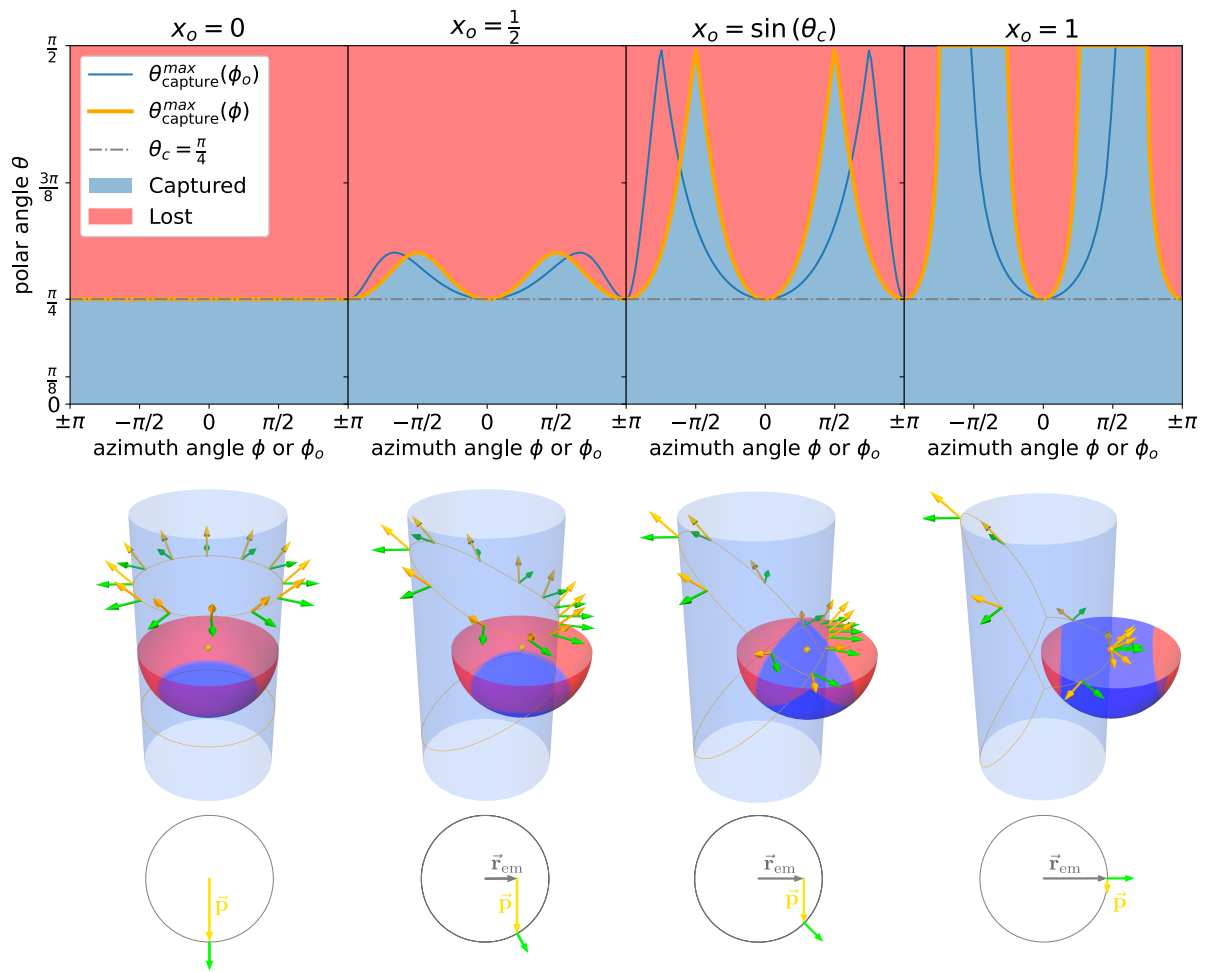}
\caption{Illustration of the solid angle under which the photons are captured as a function of the relative fraction of the radius $x_0 = {\left|\vec{r}\right|}/{R}$ at which the photons are emitted (indicated by the yellow dot). The orange (blue) line in the upper plot indicates photons that encounter the surface under the critical angle as a function of the azimuth angle $\phi$ ($\phi_0$) relative to the emission point (center of the cylinder). In the upper and middle plots the blue shaded areas indicate solid angle regions under which photon trajectories captured by total internal reflection while red shaded areas show those that are not captured. In the middle plot yellow arrows indicate directions of incident photons, while green arrows depict surface normals. In this example, the critical angle is chosen as $\theta_c = {\pi}/{4}$.\label{fig:3Dacceptance}}
\end{figure}

\begin{figure}[htb]
\begin{minipage}[t]{0.49\textwidth}
\centering
\includegraphics[width=1.0\textwidth]{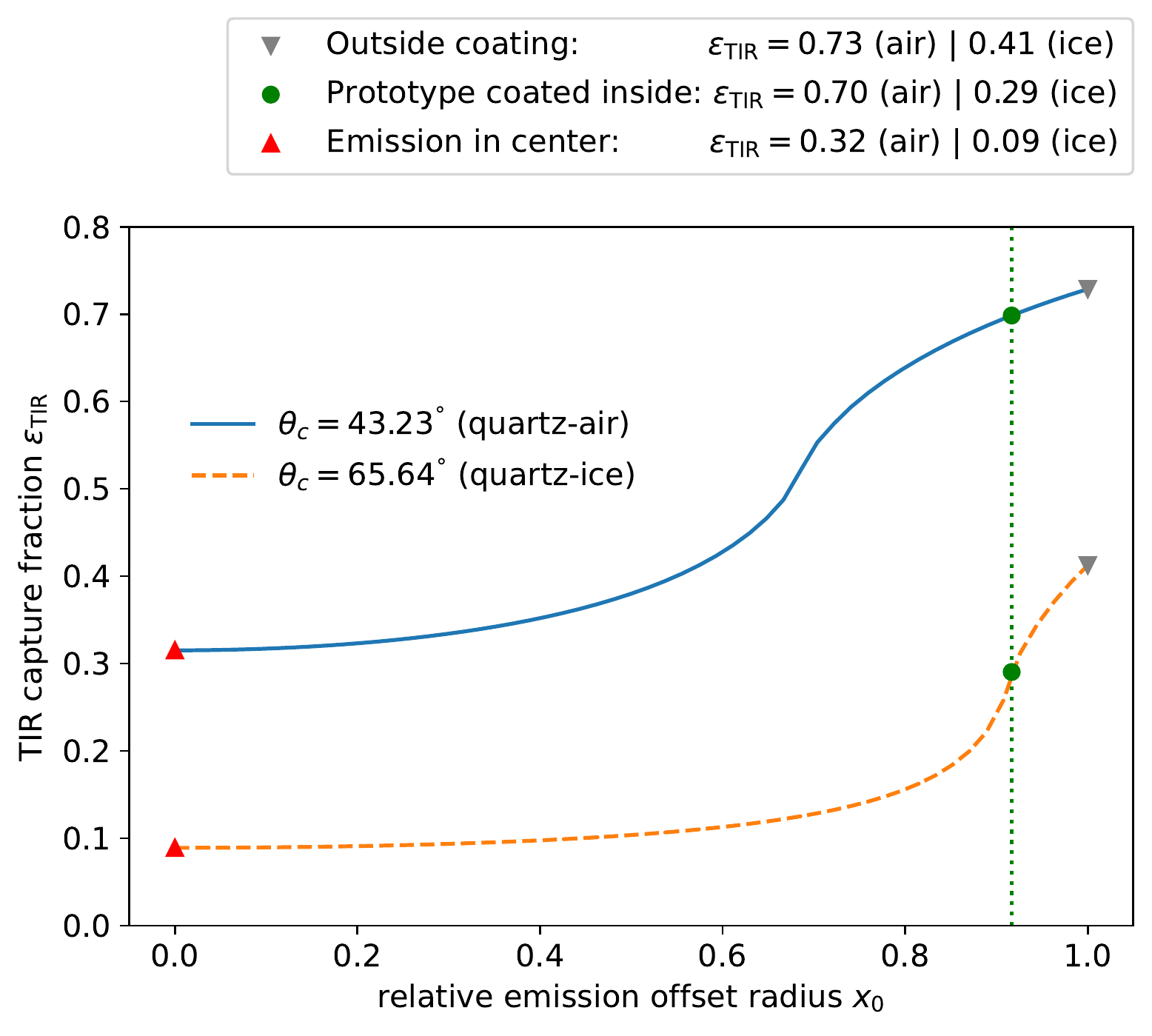}
  \caption{
  Fraction $\epsilon_{\rm TIR}$ of the solid angle under which emitted photons are captured by
  total internal reflection as a function of the offset radius \replaced{$x_0$}{$x_o$} of the
  emission point. 
  The two lines represent a module immersed in air (solid) or
  water (dashed). The green dots indicate\deleted{d} the captured efficiency for the
  prototype module, which is coated on the inside of tube. Minimal values for the
  emission point in the center of the tube \replaced{and}{as well as} maximal values for
  emission on the outside of the tube are indicated by
  triangles.\label{fig:epsilon_TIR}}
\end{minipage}
\hfill
\begin{minipage}[t]{0.49\textwidth}
  \centering
  \includegraphics[width=0.9\textwidth]{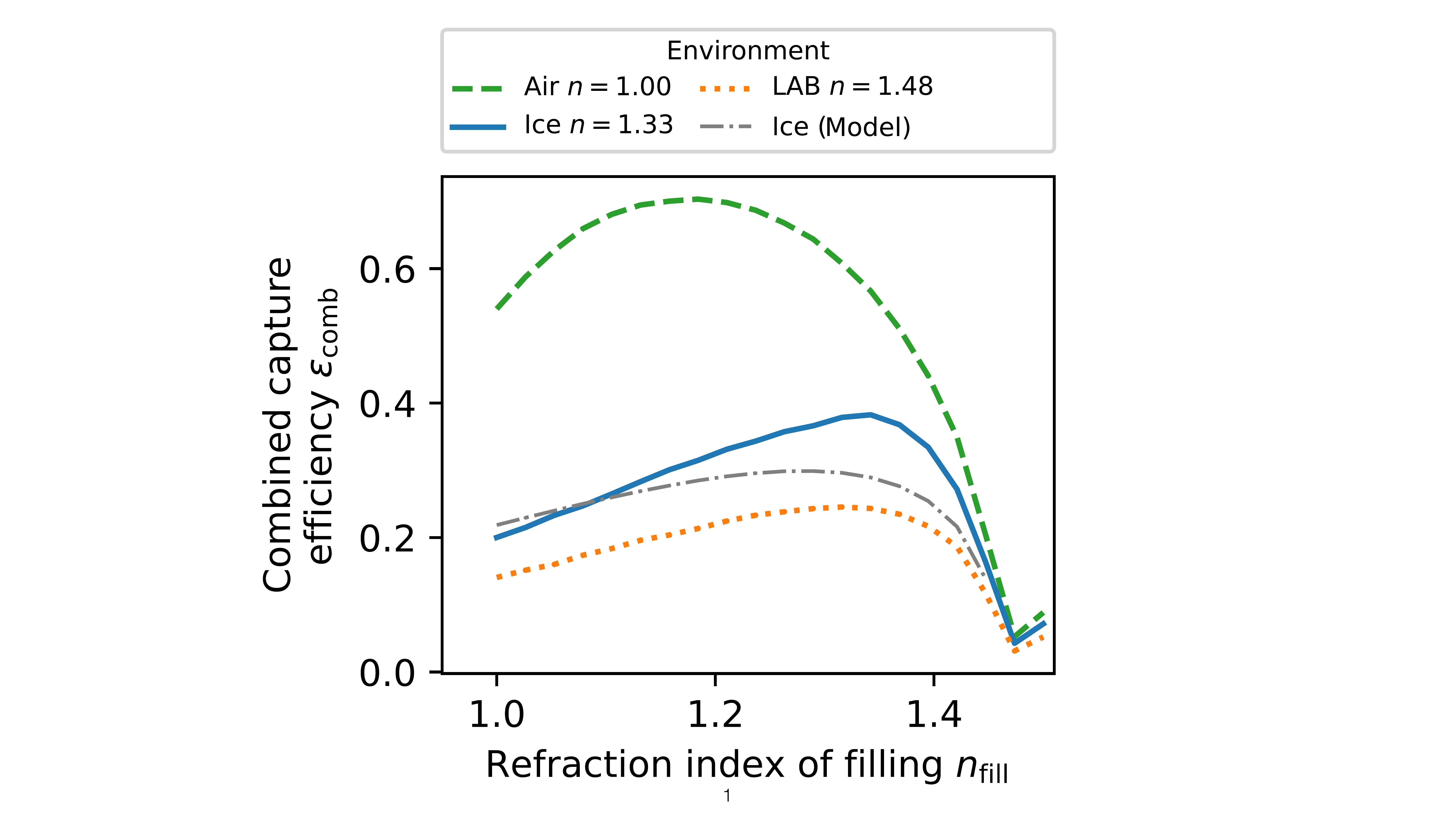}
  \caption{
  The combined efficiency for WOM operation in ice (or water), air, and LAB derived by simulating the propagation of $5\cdot 10^5$ photons \added{that are} incident isotropically. The maximum achievable efficiency is \unit[38.3]{\%} in ice at $ n\approx1.32$, \unit[70.3]{\%} in air at $ n\approx1.18$, and \unit[24.5]{\%} in LAB at $ n\approx1.32$ (refractive index for LAB taken from Ref. \cite{LAB}). \replaced{For the}{As} geometry, the prototype design is used, \added{as} specified in \autoref{sec:intro}.}
\label{fig:fig_epsilon_water_air}
\end{minipage}
\end{figure}

While the general working principle of the WOM is similar to that of a wavelength-shifting fiber, the different geometry makes a noteworthy difference in the fraction of photons that can be captured by total internal reflection. Following a purely geometrical argument, there is no difference between a hollow tube and a solid cylinder, as for any photon trajectory the incidence angle on the inner surface of a hollow tube will always be larger than the incidence angle on the outer surface. Total internal reflection in the first interaction on the outer surface is thus a fully sufficient criterion to determine whether a photon is captured and guided along the tube.
In a full cylinder the {\it no-curvature} approximation $r_{\rm tube} - r_{\rm em} \ll r_{\rm tube}$ no longer applies and \autoref{eqn:TIR} needs to be generalized to
\begin{equation}\label{eq:emissionPoint}
  \varepsilon_{\rm TIR} = \frac{1}{4\pi}\int \Theta(\left|\vec{p} \cdot \vec{n}\right| - \cos(\theta_{\rm crit})) \,d\Omega
\end{equation}
If the emission point $\vec{r}_{\rm em}$ sits in the center of the cylinder - as would approximately be the case in a wavelength-shifting fiber - this results in two symmetrical \emph{capture cones} along the symmetry axis of the fiber while most of the photons are lost even for large values of $\theta_{\rm crit}$ (left-most case in \autoref{fig:3Dacceptance}). 
As the emission point is moved outward by a fraction $x_0 = {r_{\rm em}}/{r_{\rm tube}}$, the incidence angle for photons orthogonal to the offset vector of the emission point $\vec{p} \perp \vec{r}_{\rm em}$ increases, thus increasing the overall capture fraction. 
Moving the emission point out to $x_o = \sin\left(\theta_c\right)$, even {\it horizontal} photons $\vec{p} \perp \vec{r}_{\rm em} \perp \vec{z}$ also perpendicular to the cylinder axis $\vec{z}$ subtend an angle larger than the critical angle with the surface, so that under some azimuthal angle all photons are captured independent of their polar angle. 
This region is maximized for emission on the surface, $\lim x_0 \to 1$, where again only the two cones around the surface normal vector are lost. This is the edge-case depicted in~\autoref{fig:light_trapping} close to which the WOM is operated. In \autoref{fig:epsilon_TIR} the fraction of the solid angle $\epsilon_{TIR}$ is shown under which the light is captured in the first surface interaction. For the ease of handling and to protect the coating from the environment, the coating is applied on the inside of the prototype discussed here. In this case, $x_o$ and thus the wall thickness of the tube become a relevant parameter. For the particular geometry of the prototype significant losses due to $x_0$ only occur if the module is immersed in water or ice.

\subsection{Light propagation \label{sec:flattened_model}}

Once the photons are trapped in the tube, absorption and scattering are the two mechanisms which contribute to light losses along the photon path with characteristic lengths $\lambda_{\rm abs}$ and $\lambda_{\rm scatt}$. While absorbed photons are lost directly, scattered photons may no longer be confined by total internal reflection, leading to a total attenuation characterized by the attenuation length ${\lambda_{\rm att}}$. Both attenuation mechanisms depend on the photon path length which in turn depends on the initial emission direction  
(see \autoref{fig:ray_tracing}).

Using the {\it no-curvature} approximation from above as basis, the path length distribution can be described in an analytical model -- the so-called {\it flattened model}. Therein, the tube mantle is approximated as a flat rectangle with periodic boundary conditions on the long sides.\footnote{Figuratively, the mantle is 'cut open', flattened out and photons reaching one of the long sides of the rectangle enter it again from the opposite side.}
In addition to the (polar) surface incidence angle $\theta$, the azimuthal angle $\phi$ is defined, with $\phi=0$ denoting the direction towards one end of the flattened tube (see \autoref{fig:light_trapping}). Due to the symmetry of the situation it is sufficient to consider photons that travel towards one end, limiting $\phi$ to the range $[-\pi/2, \pi/2]$. In the {\it flattened model}, the path length to the end of the tube becomes $z/(\cos(\phi)\cos(\theta))$ and the one-sided efficiency (i.e. the fraction of re-emitted photons reaching one end of the tube) is calculated as 
\begin{linenomath}
\begin{align}
  \epsilon_{\rm one-sided} (z,\lambda_{\rm att})  :&= \epsilon_{\rm TIR} \cdot \epsilon_{\rm TP}(z,\lambda_{\rm att}) \label{eqn:flattened_model1} \\
    & = \frac{1}{4\pi}\,\cdot\,\int_{-\pi/2}^{\pi/2}\,\int_{-\theta_C}^{\theta_C}\exp\left(-\frac{z}{\cos(\phi)\cos(\theta)\,\lambda_\mathrm{att}}\right) \,\cos(\theta)\, d\theta\, d\phi
    \label{eqn:flattened_model}
\end{align}
\end{linenomath}
Since $\epsilon^{\mathrm{WLS}}_{\mathrm{LY}} \approx  1 \approx \epsilon_{\mathrm{IF}}$ in \autoref{eqn:master_formula}, the one-sided efficiency becomes the predominant performance measure differentiating the WOM from the operation of a bare PMT. \added[id=AP]{The shape of this distribution can be seen in \autoref{fig_attenuation_length} where it is fit to measurements.}

To verify this simplistic model, it is compared to a \emph{Monte Carlo} (MC) simulation of the photon propagation in the tube. The GPU based implementation incorporates the full 3D geometry as well as an accurate model of absorption and scattering \cite{Thomas:2019}, see the visualization in \autoref{fig:ray_tracing}. 
The model and the simulation agree with less than < 1\% deviation for a tube length of $20\,\mathrm{mm}$ to $1200\,\mathrm{mm}$. Therefore further MC simulation, used in this work, use the model instead of the GPU simulation.

\subsection{Light transmission\label{sec:transmission}}

If the WOM is operated in a pressure housing with refractive index $n_{\mathrm{housing}}$, the transmission of light from the environment with refractive index $n_{\mathrm{env}}$ into the modules needs to be taken into account. 
In order to preserve the UV sensitivity, quartz glass can be used. The extinction coefficients is less than \unit[0.1]{\%} 
per centimeter down to a wavelength of $\lambda_{\rm inc} \sim \unit[180]{nm}$ for fused silica and down to $\lambda_{\rm inc} \sim \unit[240]{nm}$ for fused quartz \cite{transmission}, so that the dependence on the photon wavelength $\lambda_{\rm inc}$ can be neglected in most cases. 

There are two processes which compete while optimizing the refractive index of the {\it filling material} $n_{\rm fill}$ between the housing and the WOM tube:
(i) materials with low refractive index will optimize the capture efficiency $\epsilon_{\rm TIR}$ (see Equation \ref{eqn:TIR}) while 
(ii) they reduce the the efficiency of photon transmission $\epsilon_{\mathrm{TM}}$ into the module due to total internal reflection at the housing-filling material boundary.
In the co-planar model with isotropic illumination, the capture efficiency $\epsilon_\mathrm{comb}^\mathrm{co-planar}$ is derived from the Fresnel equations

\begin{linenomath}
  \begin{align}
    \begin{split}
    \epsilon_\mathrm{comb}^\mathrm{co-planar} (n_{\rm env}, n_{\rm fill}) :& = 
    \epsilon_{\mathrm{TM}}(n_{\rm env},n_{\rm housing},n_{\rm fill}) 
    \cdot \epsilon_{\mathrm{TIR}}(n_{\rm fill},n_{\rm tube}) \\
        & = \int_{0}^{\pi/2} T(n_\mathrm{env}, n_\mathrm{housing}, \theta_{\rm inc})\cdot T(n_\mathrm{housing}, n_\mathrm{fill}, \theta_j(\theta_{\rm inc})) \\
        & \qquad \quad \cdot \, T(n_\mathrm{fill}, n_\mathrm{tube}, \theta_k(\theta_j)) \sin\theta_{\rm inc}\,d\theta_{\rm inc} \\
        & \times \cos{\theta_{\mathrm{crit}}} 
    \label{eqn:epsilon_comb}
    \end{split}
  \end{align}  
\end{linenomath}

where $T(n_i, n_j, \theta_l)$ is the transmission probability at the  interface between two media of refractive indices $n_i$ and $n_j$ for an incident angle $\theta_l$ determined by Snell's law on the previous layer. 
There is an optimal value for the refractive index of the filling already in this simplified approach. However, the radii of the cylinders have a non-negligible impact here because the curved pressure housing can have a focusing or defocusing effect in the transverse plane and the inner tube could be missed altogether. 

Therefore, in order to find the optimal refractive index of the filling material for the prototype geometry, 
the cylindrical geometry of the WOM is modelled in a MC simulation%
\footnote{{In this ray tracing algorithm, written in Python, photons start from a rectangular plane which is uniformly rotated around the WOM. The photons are propagated through the WOM surfaces taking Fresnel losses and angular changes via Snell's law into account. The number of photons captured by total inner reflection are counted and divided by the total number of photons as well as the area of the generation plane to get the combined capture efficiency}.}, 
propagating incident photons with Fresnel equations and Snell's law in three dimensions. The wavelength dependence of the refractive indices is negligible and it is assumed that all refractive indices are fixed at a wavelength of $\lambda = \unit[589]{nm}$. 
The capture efficiency derived from this simulation is shown in \autoref{fig:fig_epsilon_water_air} and compared to the co-planar model. The optimal refractive index of the filling increases with larger refractive indices of the environment to a maximum at $ n\approx1.32$ with $\epsilon_{\mathrm{comb}}=38.3\%$ for water and ice.  
For water and  ice, the filling\deleted[id=AP]{s} could be ethanol \replaced[id=AP]{or}{and} methanol, which have the disadvantage that they dissolve the WLS paint, \replaced[id=AP]{or}{and} perfluoropolyether (PFPE), which is \replaced{comparatively}{comparably} expensive.


\section{Coating\label{sec:paint}}

In addition to the optimization of the WOM design based on theoretical considerations, as presented in the previous section, some components are  optimized empirically. 

The absorption probability for photons incident on a WLS paint layer is given by
\begin{linenomath}
  \begin{align}
    \begin{split}
    \epsilon^{\mathrm{WLS}}_{\mathrm{LY}} & = \left(1 - e^{-d \cdot \sigma_{\mathrm{WLS}}(\lambda)\cdot n(c)}\right) \cdot \epsilon_{\rm QY}^{\rm PL} \\
    & \simeq (1-10^{-\delta_{\mathrm{OT}}}) \cdot \epsilon_{\rm QY}^{\rm PL} \\
    & = (1-e^{-{d}/{\lambda_{\rm abs}^{\rm inc}}}) \cdot \epsilon_{\rm QY}^{\rm PL}
    \label{eq:paint_abs}
\end{split}
\end{align}
\end{linenomath}

Here, $d$ denotes the distance of a photon travelling inside the WLS matrix -- i.e. the thickness of the coating and  $\sigma_{\mathrm{WLS}}(\lambda)$ stands for the cross section of the wavelength-shifter with incident photons of wavelength $\lambda$. The number density of WLS particles along the incident photons path $n(c)$ solely depends on the concentration of WLS in the matrix $c$. The effectiveness of the light absorption can be expressed either in the form of an optical thickness $\delta_{\rm OT}(\lambda)$ or in terms of an absorption length $\lambda_{\rm abs}^{\rm inc}(\lambda) = 1/\left(\sigma_{\rm WLS}(\lambda)\cdot n \right)$. The emission efficiency of the WLS paint is denoted as  $\epsilon_{\rm QY}^{\rm PL}$.

In this chapter, the steps taken along the development of the WLS coating are outlined as well as the considerations on the coupling of tube and PMT. First the test setup is presented which was build to aid the optimization as well as to measure the WOM characteristics presented in the next section. 

\subsection{Test stand \label{subsection_experimental_setup}}

\begin{figure}[H]
\includegraphics[width=\textwidth]{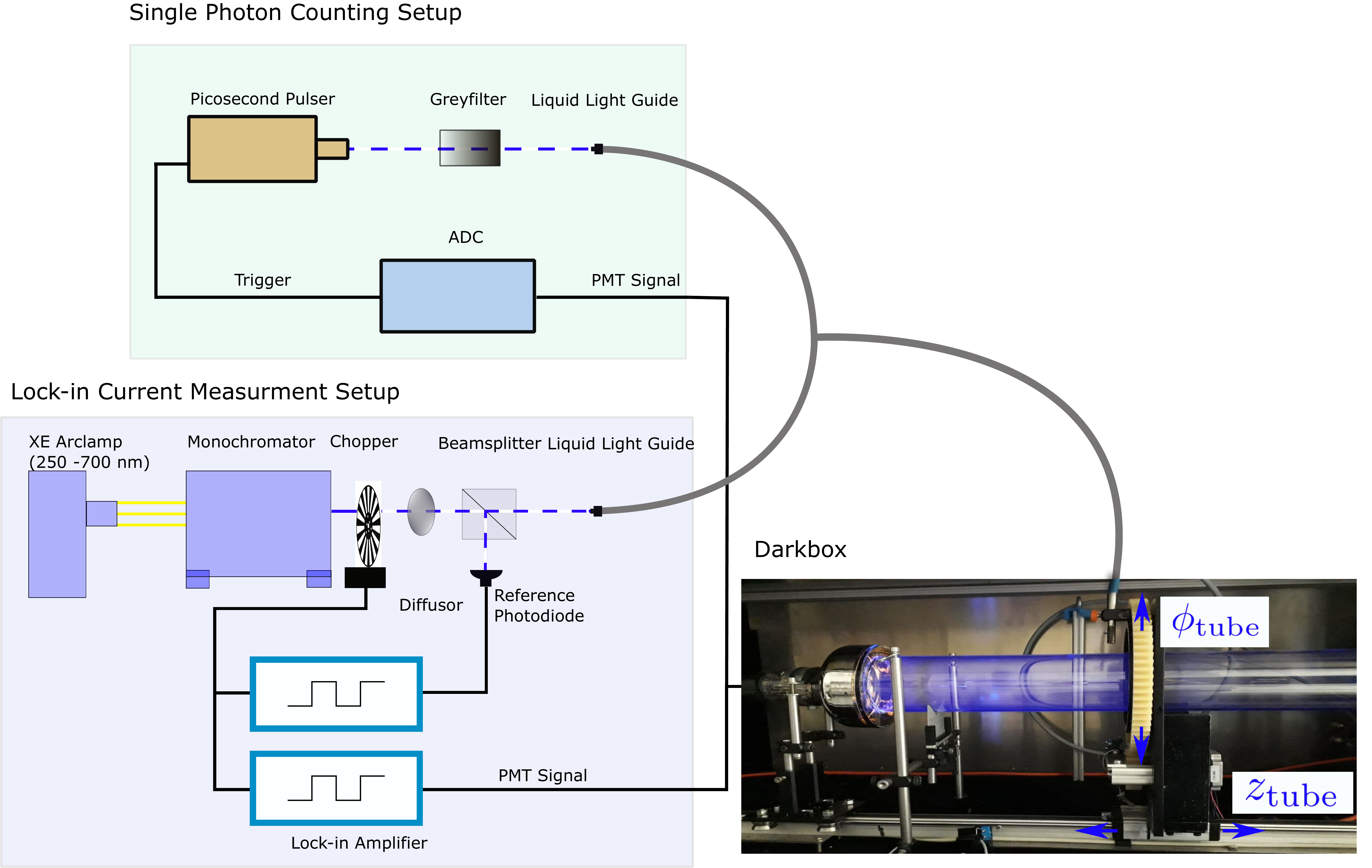}
\caption{Schematic of the WOM test stand{ with a photograph of the darkbox.}}
\label{fig_teststand_Mainz}
\end{figure}

The scope of the test stand is to measure the local sensitivity of a WLS tube and its timing characteristics. The setup is schematically depicted in \autoref{fig_teststand_Mainz}. 
In the test stand, a Xenon arc lamp \added{\cite{QuantumDesign}}  
serves as light source for wavelengths between \SIrange{250}{700}{\nm}. The lamp is coupled to a monochromator \added{\cite{MSH300}} of which the slit width is set in a way that the bandwidth of the selected wavelength is $\sigma_{\lambda} = \SI{1.06}{\nm}$.

The monochromatic beam is chopped to allow for a signal read out via \emph{lock-in} amplifier \added{\cite{zurich}}. The chopper frequency is fed into two lock-in amplifiers which demodulate a PMT signal. 
Subsequently the light is passing a diffusor which shapes the beam to a homogeneous Gaussian profile. 
The beam is split up by a beam splitter. The reflected fraction of the beam illuminates a reference photodiode \added{\cite{Photodiode}} to correct for lamp intensity variations, while the transmitted fraction is guided into a dark box to illuminate an WLS-tube using a liquid light guide \added{\cite{Thorlabs}}. 
The end of the light guide in the dark box can be moved along the symmetry axis (the $z_{\mathrm{Tube}}$ position) and the azimuthal angle $\phi_{\mathrm{Tube}}$\footnote{Note that this is a different definition than the  the variable $\phi$ in the flattened model in \autoref{eqn:flattened_model}}. The size of the illumination spot is around \SI{1}{\cm} in diameter.   
Thereby, selected points in $\phi_{\mathrm{Tube}}$ and $z_{\mathrm{Tube}}$ on the tube surface can be illuminated to determine the local efficiency. The light is detected at both tube ends using PMTs \cite{hamamatsu}. The PMT surface is described in polar coordinates with the radius $r$ and the angle $\psi_{\mathrm{PMT}}$.   

In this \emph{lock-in} setup, the efficiency of the WOM for a given light source position and incident wavelength is calculated as 
\begin{linenomath}
\begin{align}
    \epsilon_\mathrm{lock-in}(\lambda, \phi_{\mathrm{Tube}}, z_{\mathrm{Tube}}) = \frac{I^\mathrm{corr}_\mathrm{sig}(\lambda, \phi_{\mathrm{Tube}}, z_{\mathrm{Tube}})}{\frac{1}{N_{\rm steps}}\sum_{i} I^\mathrm{corr}_\mathrm{ref}(\lambda, \psi_{\mathrm{PMT,i}})}\qquad
    \label{eq:eps_meas}
\end{align}
\end{linenomath}
Here $I^\mathrm{corr}_\mathrm{sig}$ refers to the PMT output current {corrected by the reference diode signal}. For the reference measurement, the light source illuminates the PMT directly. The current $I^\mathrm{corr}_\mathrm{ref}$ is measured on the tube radius by varying $\psi_{\rm PMT}$ in a total of $N_{\rm steps}$ steps around the circumference, in our case $N_{\rm steps} = 36$. The mean of these measurements is calculated to take the PMT sensitivity variations on the photocathode into account. The resulting currents of signal and reference measurement are corrected for the wavelength dependent quantum efficiency of the PMT as well as for intensity variations of the light source. 

An overview of systematic uncertainties on the efficiency determination in the test stand is given in Ref. \cite{Rack-Helleis:2019}. The largest uncertainty is the coupling of the WOM tube to the PMT using optical gel instead of a gel disc. An overall systematic error of approximately \SI{4}{\%} on the efficiency determination for the prototype tube is obtained. 
{This value was obtained by repeatedly coupling the tube and PMT with gel to measure the variance of the one sided efficiency induced by the losses at the interface $\epsilon_{\rm IF}$. Additionally the variation of the paint layer is taken into account \added[id=AP]{(compare \autoref{fig_efficiency_2D})}.}
The values for the currents $I^\mathrm{corr}_\mathrm{sig}$ and $I^\mathrm{corr}_\mathrm{ref}$ are averaged over a time window of $2\,\mathrm{sec}$ at a chopping frequency of $120\,\mathrm{Hz}$ and the standard deviation is used as uncertainty. 

The described setup can be adjusted for \emph{single photon} read-out by  replacing Xe-Lamp, monochromator and chopper with a \SI{375}{\nm} pico-second pulser \cite{Rongen_2018}.
The pulser is dimmed so that \SI{\sim 6}{\%} of the waveforms recorded with a triggered fast ADC \cite{teledyne} include one photon signal within the expected time window.
For the \emph{single photon} read-out, the collection efficiency of the tube $\epsilon_{\mathrm{single}}$ is determined by the ratio:

\begin{equation}
    \epsilon_{\mathrm{single}} = \frac{n_{\mathrm{sig}}(\lambda, \phi_{\mathrm{Tube}}, z_{\mathrm{Tube}}) - n_{\mathrm{bkg}}}{\frac{1}{\mathrm{N_{\rm steps}}}\sum_{\mathrm{i}} \left( n_\mathrm{ref}(\lambda, \psi_{\mathrm{PMT,i}}) -n_{\mathrm{bkg}}\right)}
    \label{eq:single_eps}
\end{equation}

where $n_{\mathrm{sig}}$ denotes the number of photons detected when illuminating the point $(\phi_{\mathrm{Tube}}, z_{\mathrm{Tube}})$ with wavelength $\lambda$. 
The number of photons detected when directly illuminating the PMT is denoted by $n_{\mathrm{ref}}$ and $n_{\mathrm{bkg}}$ is the number of photons falsely reconstructed from background light. 
The stability of the pulser has been measured over a \SI{12}{\hour} cycle, and the light output variance was found to be smaller than \SI{2}{\%}, {thus a subdominant contribution to the overall systematic error }. The statistical error is calculated based on the number of measured photons $n$ for each measurement. 
Since the same systematics as in the lock-in amplifier setup apply, the estimated systematic error for this setup amounts to approximately \SI{4}{\%}.

\subsection{Chemical composition\label{section_paint}}

Wavelength-shifting is a special case of photofluorescence. The WLS molecules are excited when absorbing short wavelength photons. After a decay time on the order of \SIrange{1}{2}{\ns} \cite{wls_overview} depending on the specific wavelength-shifter, the molecule returns to the ground state and a photon of larger wavelength is emitted. A fraction of the energy is dissipated non-radiatively.

The WLS paint for the WOM is selected from a large number of variants with regard to the following criteria as treated in detail in Ref. \cite{Hebecker:2014}:

\begin{itemize}
\setlength{\itemsep}{0pt}
\item maximal overlap of the emission spectrum and the sensitivity of the readout PMT 
\item large Stokes shift, i.e. minimal overlap of the absorption and emission spectrum
\item maximal transparency of the WLS paint for re-emitted photons
\item similar refractive index of the coating and the WLS tube material
\item good adhesiveness and mechanical properties
\item optical thickness to enable reaching high concentration of WLS film to absorb 100\% of the light for a broad spectrum
\end{itemize}

Best results are obtained with a solution of Toluene containing \unit[213]{g/l} of the plastic po\-ly\-ethyl\-methyl\-acrylate (PEMA) \cite{paraloid}, which provides the matrix for the wave\-length-shift\-ing mole\-cules, and a wave\-length-shift\-ing dye mix of \unit[1.3]{g/l} Bis-MSB and \unit[2.6]{g/l} p-Terphenyl per liter\footnote{The proportions given apply to the mixing process of the paint. In the paint layer of a coated tube, proportions differ as the toluene evaporates during drying.}. The absorption spectrum of p-Terphenyl lies well below the Bis-MSB absorption spectrum. Since the emission spectrum of p-Terphenyl lies within the absorption spectrum of Bis-MSB, adding p-Terphenyl yields a sensitivity enhancement in the lower UV region.


\subsection{Coating process\label{sec:coating}}

For  the  tube  material, PMMA  (Polymethylmethacrylate) and quartz glass are considered due to their optical properties with attenuation lengths for photons of multiple metres \cite{Beise:2019, transmission}. 
PMMA and PEMA are both soluble in toluene, which yields a transit region in which the two materials are mixed. This guaranties a strong bond, able to withstand shearing forces generated by the different coefficients of thermal expansion over a wide temperature range. Since quartz glass is not soluble in toluene, the paint is bonded by \added{van der Waals} forces only.

Since the bond between quartz glass and the paint is purely  of \added{van der Waals} nature, any contamination on the tube surface affects the bonding strength of the paint. When the tube surface is treated with a sequence of citric acid, acetone and isopropanol, we find good adhesion in the subsequent coating. An alternative method is the usage of caustic soda and cleaning agent  
\cite{mucasol}. Both methods yield satisfying results. Current versions of coated quartz tubes have proven to be stable over time. We found no signs of delamination in several freezing cycles. Additionally, no significant deterioration of the paint layer when exposed to laboratory light was observed.

The coating can be applied to the inside or outside of the cylinder. It is also possible to coat both sides, but generally high enough absorptivity for the incident UV light can be obtained with a single coating. Some major differences between in and outside coating are summarized as follows:
 \begin{itemize}
     \item  Coating on the inside shifts the light emission point towards the center, resulting in a reduced capture efficiency, as discussed in \autoref{sec:TIR}. The performance loss will increase with the thickness of the tube and the refractive index of the environment in which the coated tube is deployed.
     \item While quartz glass has high transmission down to wavelengths of \SI{180}{\nm}, PMMA is generally opaque to light below \SI{300}{\nm}~\cite{Abdel-Mottaleb:2009vo} but in commercial products it is often doped with additional UV-absorbers to reduce aging in sun light, thus limiting the UV light yield. Quartz glass can therefore be coated on either side, while PMMA performs best when coated outside.
     \item Coating a tube on the inside allows for easier handling. Impurities such as fat or dirt act as scattering centers for photons travelling inside the tube. Under UV light illumination, contaminated areas are clearly visible presumably because photons couple out of the tube at these sites. The WLS paint is hydrophilic and therefore delaminates as a whole when immersed in water rendering the tube opaque to light.  
     \item In an assembly, an inside coated tube will not interact chemically with the medium surrounding it, allowing for the surrounding medium to be chosen freely. While embedding the module in a housing can alleviate the problem in a similar way for outside coated tubes, a filling material is in general required between the housing and the tube~(see \autoref{fig:fig_epsilon_water_air}), transferring the problem to chemical compatibility with the filling material.   
 \end{itemize}
 
\begin{figure}[H]
\includegraphics[width=\textwidth]{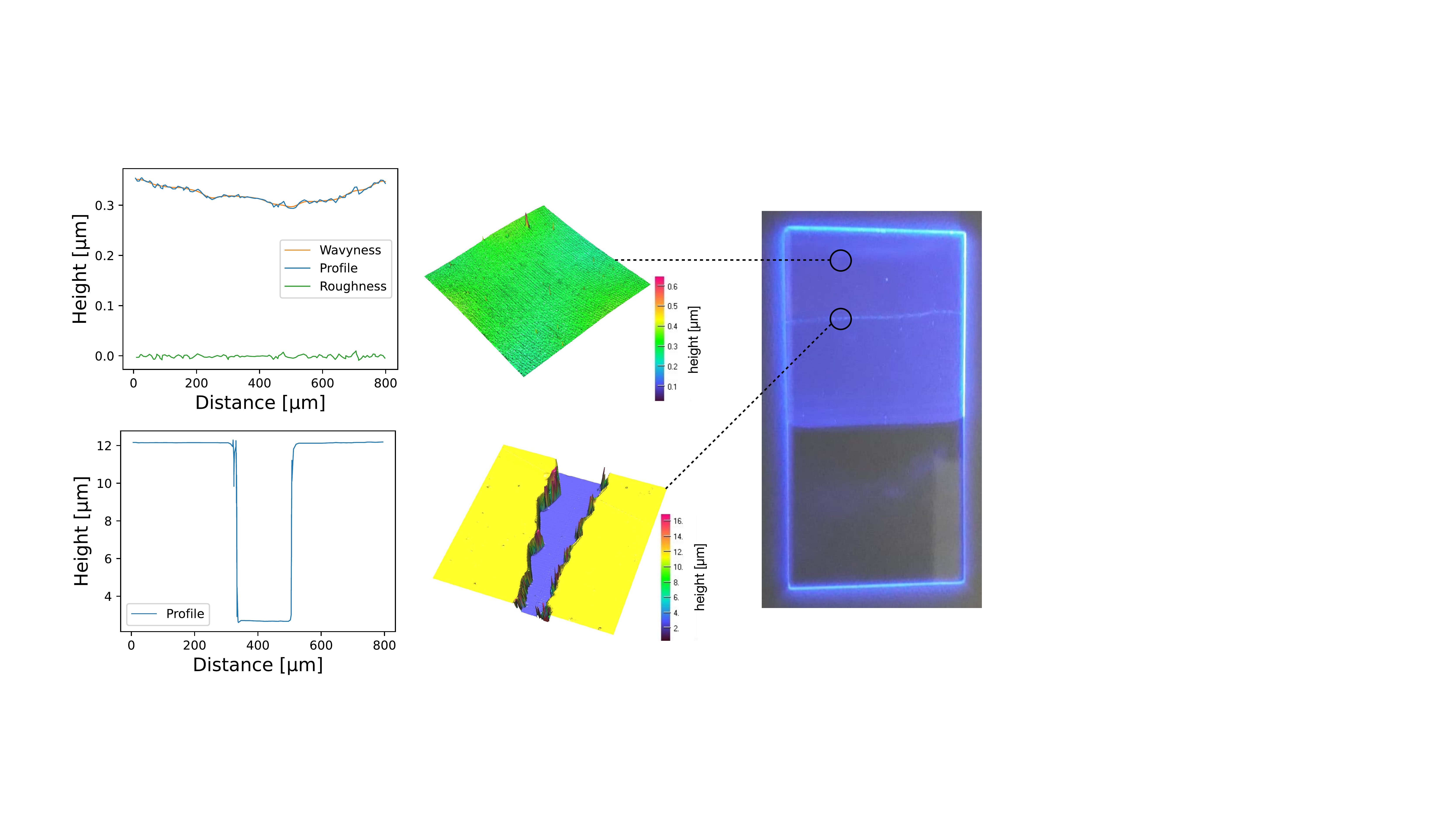}
\caption{{Profilometer measurements of two different paths on a slide, coated with the wavelength shifter, are shown. In each row the measured profile is shown on the left and a 3 dimensional profile is shown in the middle. The path in the bottom row comprises a scratch in order to measure the thickness of the coating. A photograph of this slide is shown on the far right in which the approximate measurement points are indicated.}}
\label{fig:profilometer}
\end{figure}

In order to apply coatings to tubes, two simple techniques have been developed, which are both based on the industrially employed \textit{dip-coating} method~\cite{Brinker2013,RIO2017100}. For an external coating,  the tube is vertically immersed in the paint and removed at a controlled speed, to yield a homogeneous coating of adjustable thickness. Plugs are used to keep the inside of the tube paint free. In order to achieve an internal coating, the tube is  filled completely with paint which is then released at a constant flow rate using a valve. Due to the viscosity of the WLS solution, a higher velocity of extracting the paint in the coating process leads to a thicker paint layer~(see~\autoref{eqn:coating_thickness}). Both coating procedures are performed at a temperature of \SI{20 \pm 1}{\celsius} and the tubes stay immersed in the paint for \SI{90}{\s} before paint or tube are extracted at a fixed velocity.
Both processes are operated in the Landau-Levich-Derjaguin (LLD) regime~\cite{Derjaguin:43d} where for sufficiently high viscosity $\eta$ and coating speed $v_{\rm coating}$ the wet-film thickness is given as~\cite{Brinker2013,RIO2017100} 

\begin{equation}
    d_0 = 0.8\cdot \sqrt{\frac{v_{\rm coating}\eta}{\rho g}} \label{eqn:coating_thickness}
\end{equation}

Here $\rho$ denotes the density of the paint and $g$ is the gravitational acceleration. As the solvent evaporates, a dry film of constant thickness $ d =\epsilon_{d} \cdot d_0$, with $\epsilon_{d}= {\rho_{\rm PEMA}}/{\rho_{\rm paint}} = 0.198$ is deployed on the surface. The relative factor $\epsilon_{d}$ corrects for the paints density before and after coating. Besides its simplicity, it is found that these methods yields excellent surface quality. Using profilometer measurements on microscope slides coated with this technique, a small-scale surface roughness of $< 5\,{\rm nm}$ and a wavyness of $50-100\,{\rm nm}$ on scales of $10\,\mu \mathrm{m}$ was found, see \autoref{fig:profilometer}. 

\begin{figure}[htb]
\begin{minipage}[t]{0.49\textwidth}
\centering
\includegraphics[scale=0.55]{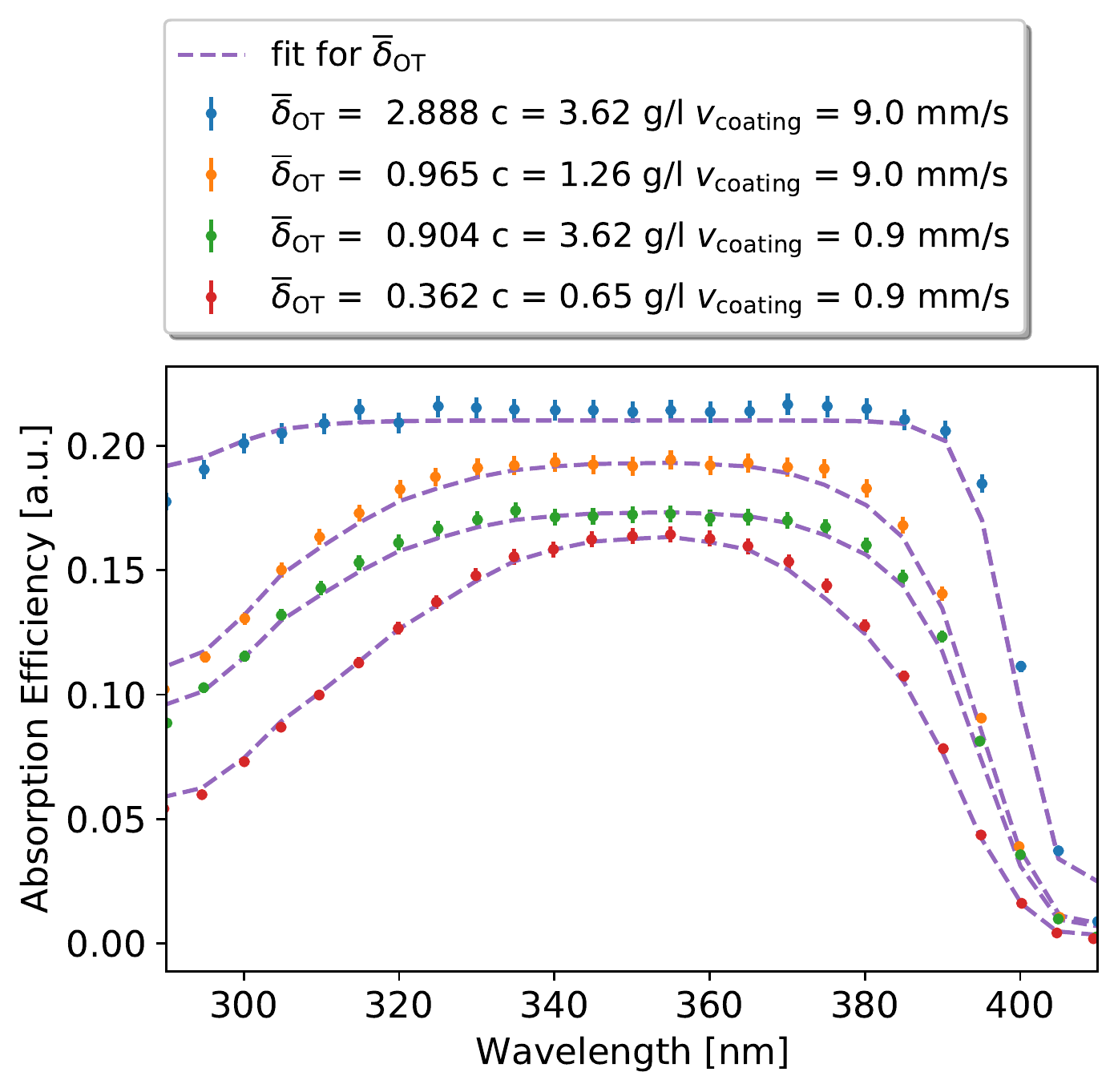}
\caption{Absorption efficiency for different coating velocities $v_{\mathrm{coating}}$ and concentrations $c$ of Bis-MSB in a WLS paint mixture of $400\,\mathrm{ml}$ anisole and $85.1\,\mathrm{g}$ PEMA (the second WLS was not used here). {The methodology of the experiment is described in Ref. \cite{Hebecker:2014}.} Shown in purple are fits to the different absorption efficiencies using \autoref{eq:paint_abs} multiplied by a normalization constant. }
\label{wls_thickness_efficiency}
\end{minipage}
\hfill
\begin{minipage}[t]{0.49\textwidth}
\centering
\includegraphics[scale=0.55]{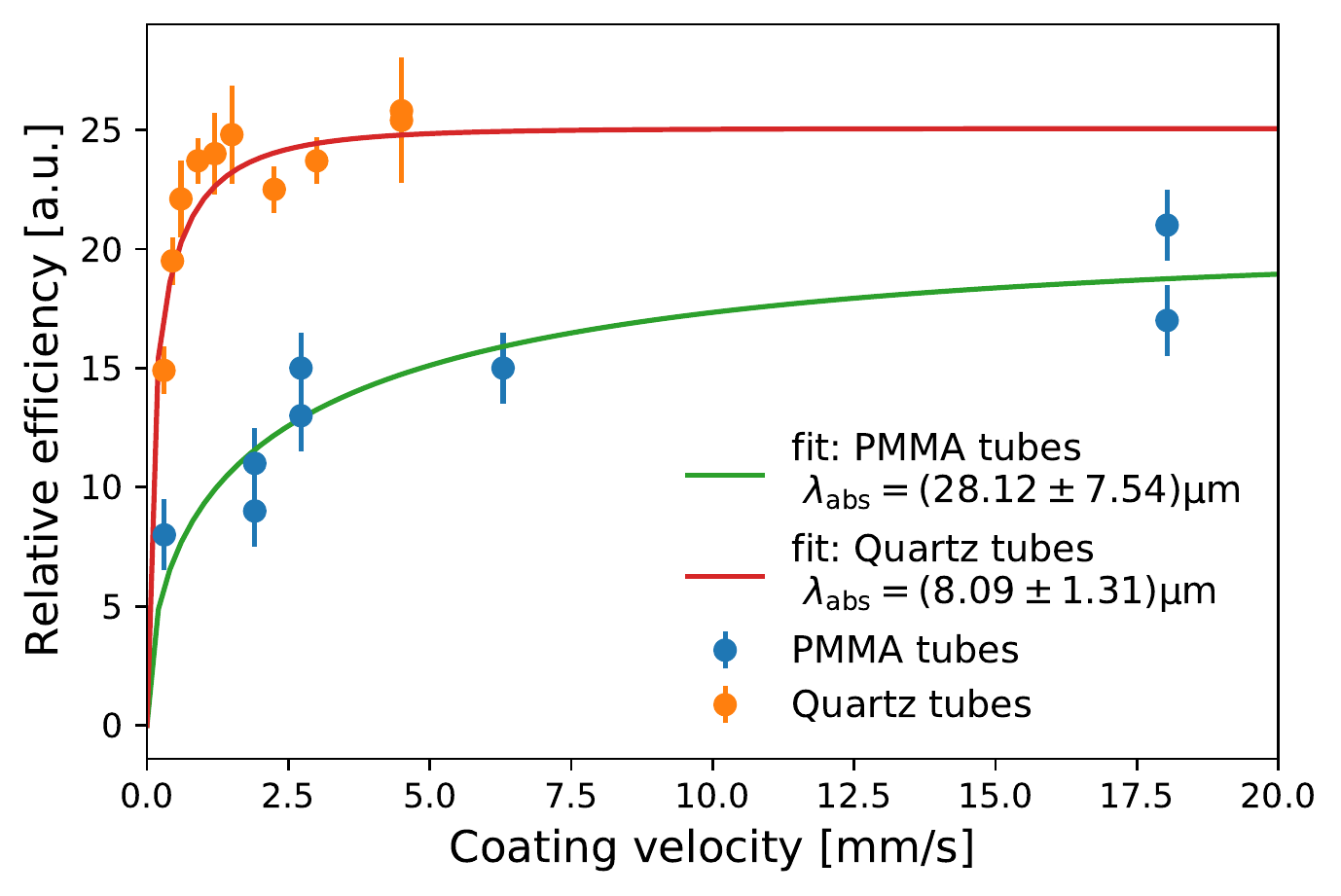}
\caption{One-sided efficiency at \SI{375}{nm} versus the coating speed for quartz tubes and PMMA tubes. Efficiencies are read out by a PMT \unit[150]{mm} away from the illumination point. The solid lines show a fit with the thickness and absorption model from Equations \ref{eq:paint_abs} and \ref{eqn:coating_thickness}.}
\label{fig_efficiency_combined_coating_speed}
\end{minipage}
\end{figure}

In \autoref{wls_thickness_efficiency} the effect of the variation of coating velocity $v_{\textrm{coating}}$ and concentration of the wavelength-shifter Bis-MBS $c_{\textrm{Bis-MBS}}$ on the absorption of the paint layer as a function of the wavelength is shown. To verify, that the absorption efficiency follows \autoref{eq:paint_abs}, we use the $\sigma_{\mathrm{WLS}}(\lambda)$ dependency from the lowest absorption curve to estimate the optical depth for the other coating speeds and concentrations. Fit and data values are in good agreement.  
Comparing the different variations, it can be observed that high paint layer thicknesses lead to an overall broadening of the absorption spectrum up to a maximum where all incident photons are absorbed. In this way, the absorptive properties of the WLS paint can be optimized.

In \autoref{fig_efficiency_combined_coating_speed}, the relative light yield of PMTs at the end faces of quartz and PMMA tubes coated on the outside is shown as a function of the coating velocity. The incident light has a wavelength of \SI{375}{\nm}. For quartz the same tube is coated, measured and the coating removed again. This procedure is repeated at different velocities and twice for the highest velocity. For PMMA tubes the coating can not be removed, so a different individual PMMA tube is coated for each measurement. As soon as nearly complete absorption of the injected light is achieved, a further increase in the coating thickness yields no further increase in efficiency. 
For the highest velocity, the two efficiencies obtained in the two coatings of quartz glass differ by significantly less than the measurement error, indicating that the coating process is well reproducible. In comparison, the results achieved for PMMA tubes show a larger spread and significantly lower efficiency values (see \autoref{fig_efficiency_combined_coating_speed}), indicating a less reliable coating process. The larger value simultaneously obtained for the absorption length for incident light $\lambda^{\rm inc}_{\rm abs}$ possibly suggests that the tube surface is partly dissolved by the toluene in the coating process, thus locally diluting the wavelength-shifter concentration.\\
The obtained absorption length $\lambda^{\rm inc}_{\rm abs}$ for quartz can be transformed into an optical depth using \autoref{eq:paint_abs} and the estimated paint thickness of \added[id=AP]{$2 \times 17\,$\SI{}{\micro\metre}} (paint layer on both sides of the tube contribute) to be $\delta_{\mathrm{OT}} = 1.83 \pm 0.30 $. This number lies well within the different optical thicknesses obtained in \autoref{wls_thickness_efficiency}.\\ 

\begin{figure}[htb]
\begin{minipage}[t]{0.49\textwidth}
\centering
\includegraphics[width=1.05\textwidth]{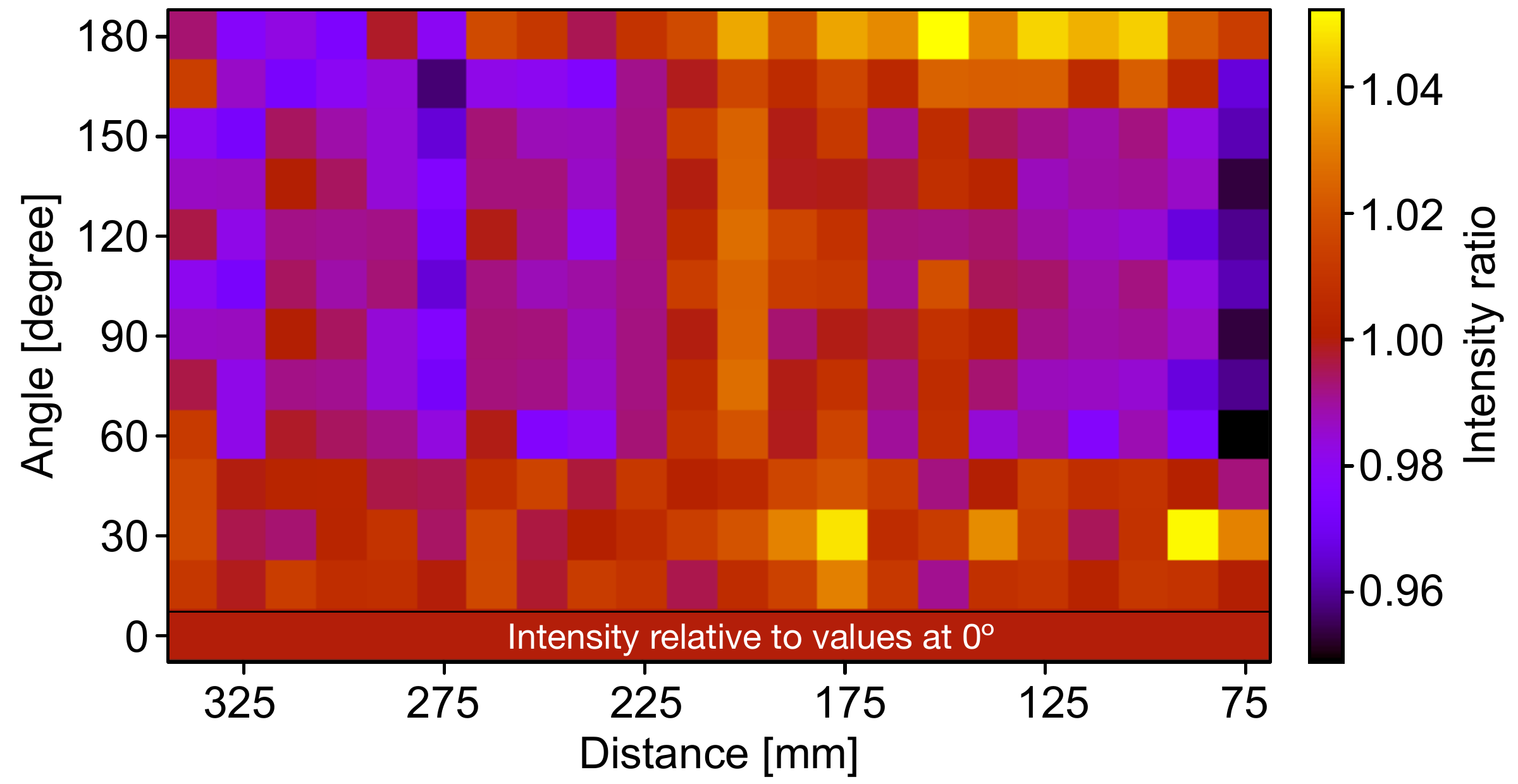}
\caption{A 2D-scan of the relative efficiency of a quartz tube. For a given longitudinal position, the efficiency is relative to the corresponding value at $0^\circ$, in order to correct for the longitudinal intensity dependence. The angular variation of the efficiency is less than $\pm\unit[5]{\%}$.}
\label{fig_efficiency_2D}
\end{minipage}
\hfill
\begin{minipage}[t]{0.49\textwidth}
\centering
\includegraphics[width=0.94\textwidth]{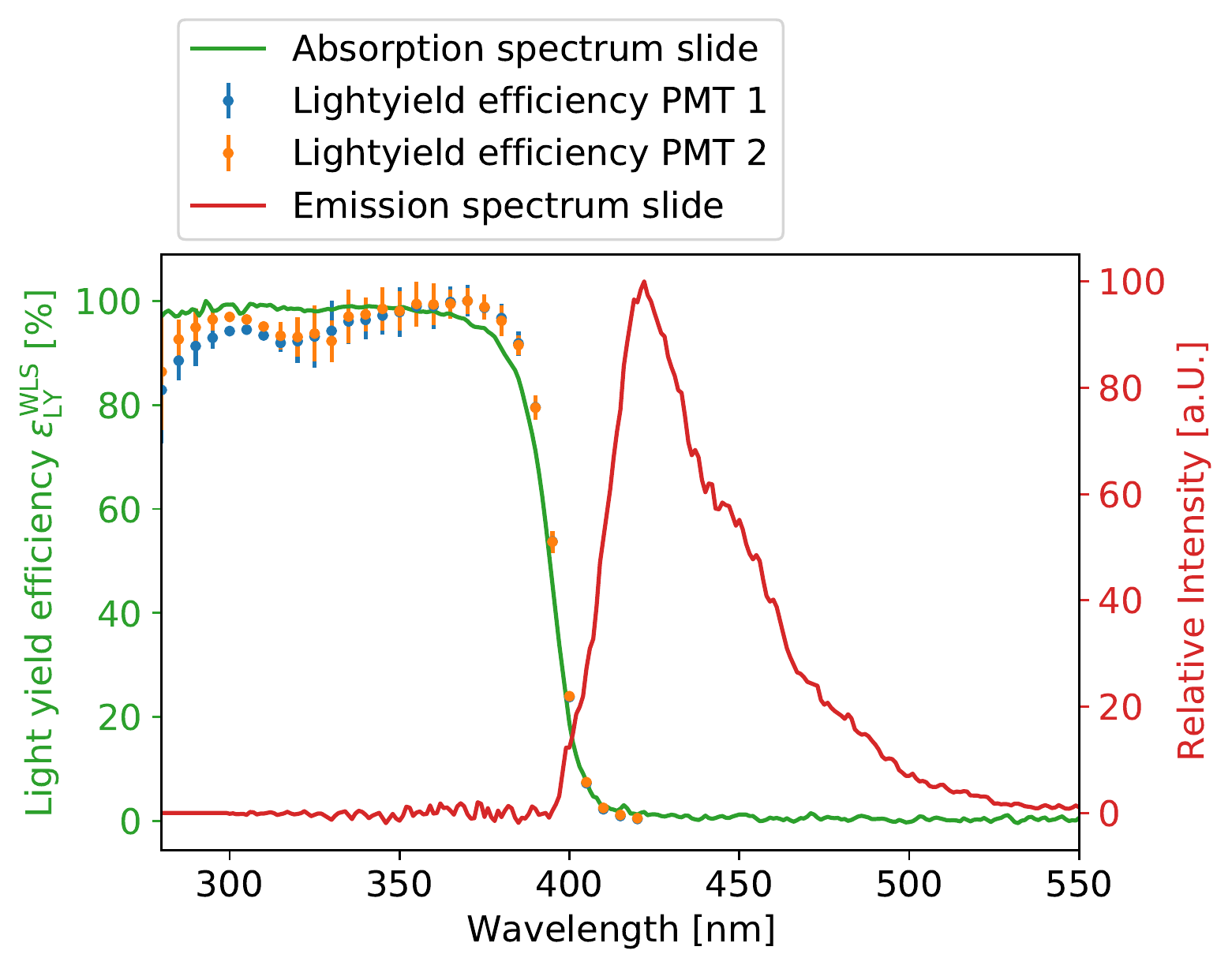}
\caption{Absorption (green) and emission spectrum (red) of a WLS-coated quartz slide. The light yield efficiency $\epsilon_{\mathrm{LY}}^{\mathrm{WLS}}$ measured on the prototype tube is shown in blue and orange. {The methodologies of the experiments are described in Refs. \cite{Binn:2018, Hebecker:2014}}.}
\label{fig_absorption_emission_wls}
\end{minipage}
\end{figure}

In \autoref{fig_efficiency_2D} a measured relative efficiency over half the surface of a quartz tube is shown. The current $I_{\mathrm{sig}}(z,\phi)$ is measured  in steps of $10^\circ$ in azimuth angle $\phi$ and \SI{10}{\mm} in cylinder height $z$. The intensity value at a given position $z$ is divided by the corresponding value at $\phi = 0$. Concluding an overall relative variation of the paint layers efficiency of \SI{\pm 5}{\%} was obtained.

\subsection{Absorption and emission\label{sec:paint_abs_emi}}

In order to measure the properties of the WLS coating, quartz microscope slides were coated with a thin film of WLS paint. The coated slides were then illuminated by a \unit[365]{nm} UV LED focused on the paint layer. The angular distribution of the emission was measured in the UV band ($\lambda$ = \unit[365 $\pm$ 10]{nm}) to test for a complete absorption of the UV light, as well as in the optical range ($\lambda$ = \unit[450 $\pm$ 40]{nm}) to investigate the re-emission. In \autoref{fig_absorption_emission_wls} the measured absorption and emission spectrum of the WLS paint are shown together with an absorption measurement of a fully coated quartz tube. 

Between \SI{280}{\nm} and \SI{400}{\nm}, the paint is around \SI{92.7}{\%} absorptive for a sufficiently thick paint layer. The overlap between absorption and emission spectrum is a small band between \SIrange{400}{420}{\nm}. Therefore, by calculating the overlap, a re-emission and consecutive capture probability of \SI{99.61}{\%} is expected in air. The emission spectrum peaks around \SI{440}{\nm}{, which is} the high quantum efficiency region for standard PMTs {that have} a borosilicate window and a bialkali photocathode.



 \subsection{Deterioration}
 
The setup described in \autoref{sec:paint_abs_emi} was further used to explore possible deterioration of the paint layer due to high UV exposure. It is found that focusing the light output of the diode with \unit[100]{mW} on a spot size of \unit[1]{mm$^2$} over several hours leads to a deterioration of the paint layer. A diffuse illumination of the entire slide with \unit[1250]{mm$^2$} using the same light intensity shows no damaging effect on a time scale of days. 
\replaced[id=AP]{In conclusion, }{Concluding} \replaced[id=AP]{with a damage threshold on the order of mW/mm$^2\times$h deterioration of the paint due to illumination is not a limiting factor.}{the deterioration of the paint due to illumination is not a limiting factor, since the damage threshold is on the order of mW/mm$^2\times$h.} 
The UV light intensity in detector experiments, in which the WOM is a suitable light sensor, will be orders of magnitude lower.

Neither degradation of efficiency with time nor after exposing the tubes to cold temperatures were found.

\subsection{Optical coupling}
\label{sec:optical_couping}

In order to minimize photon losses at the interface between end face of the WOM tube and PMT glass, different methods of optically coupling the two components  are used. For repeatability of different experiments, an optical coupling gel \cite{eljen} was used in most experiments. However for a deployable WOM, where a disassembly of the module is not desired, optical glue \cite{norland2} is the preferred choice. The refractive index of the cured glue is chosen to lie between quartz glass ($n = 1.46$) and borosilicate glass ($n=1.51$, used for the PMT surface) to minimize Fresnel losses at the interface. It should be noted that flat surface PMTs are preferable when using optical glue, since spherical PMTs require chamfered ends of the WLS tubes to fit the curvature of the PMT perfectly. If this is not the case, cracks in the glue layer are observed which likely reduce bond strength and transmission efficiency of the glue. Given the transmission from the datasheet and the Fresnel losses at the interface (calculated using the angular distribution of the light output of the WOM from simulation), a transmission probability of $\epsilon_{\rm IF}$\SI{>90}{\%} is expected when using optical glue for the coupling. 

It would be desirable to concentrate the light emission of the WOM from the tube cross-section onto a compact area,  allowing to further reduce the photocathode area of the readout PMT. While conservation of etendue  does not prohibit this if the cross-sectional area is preserved, a viable solution was not found. Starting from theoretical calculations~\cite{Falke:2014}, several approaches have been investigated in simulation~\cite{Thomas:2019,Schnur:2020} and experiment~\cite{Hebecker:2021}. It is found that all lead to severe efficiency losses\footnote{In simulation this is mostly due to photons being reflected back into the tube} and are therefore not suitable for the WOM design.



\section{Characterization \label{sec:efficiency_measurement}} 

In order to compare the WOM with other light sensors, measurements of its characteristics are conducted, i.e. the one-sided efficiency $\epsilon_{\rm one-sided}$, see  \autoref{eqn:flattened_model}, the transit time spread, and the dark noise. The effective area and the signal-to-noise ratio (SNR) are derived from these values as well as the efficiencies discussed in the previous sections.


\subsection{Efficiency}

\begin{figure}[htb]
\begin{minipage}[t]{0.47\linewidth}
\centering
\includegraphics[width=0.99\linewidth]{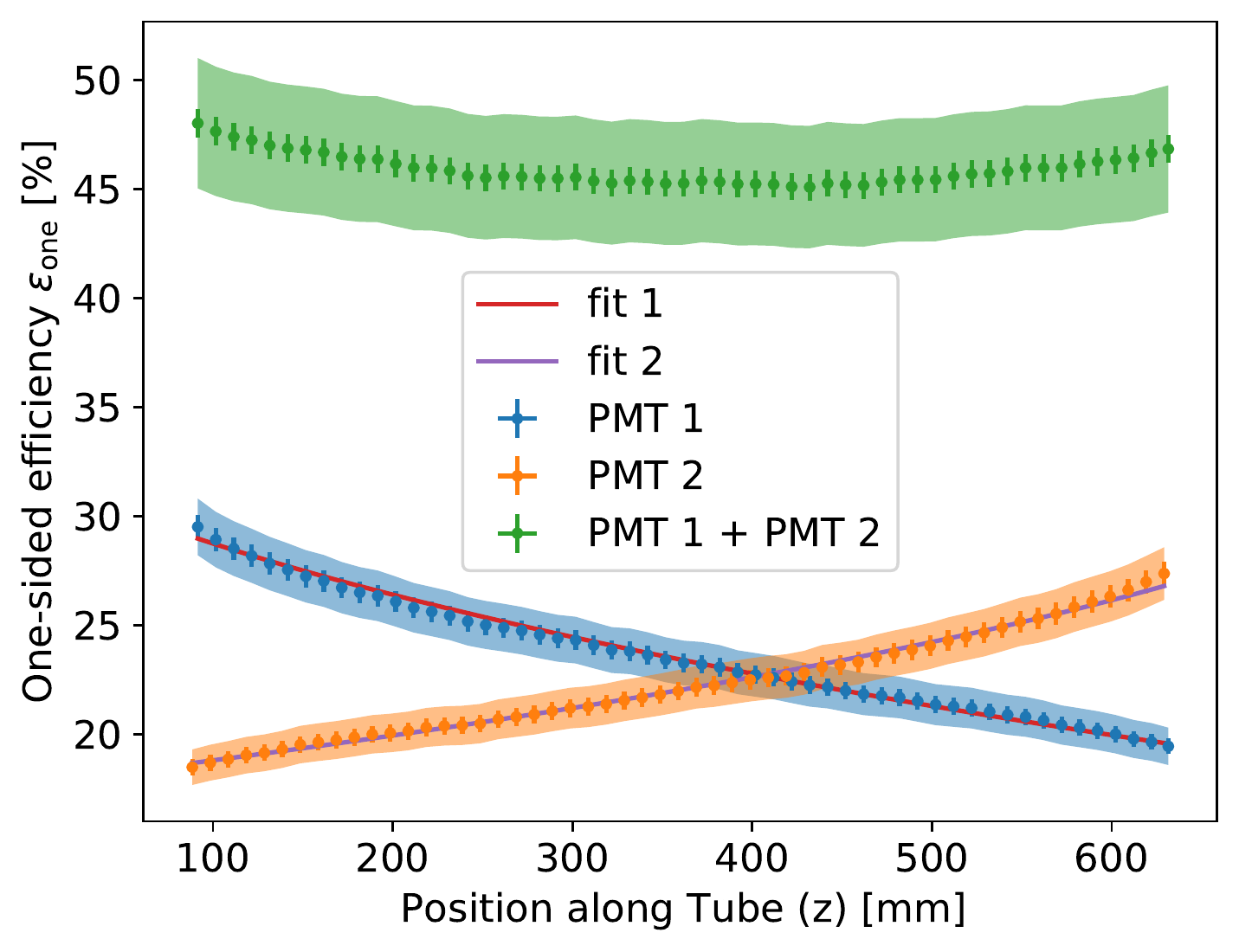}
\caption{Efficiency as a function of the distance from the ends of the tube for a quartz tube \cite{hsq300}. The flattened model efficiency is used as fit function to extract attenuation lengths. The injected light has a wavelength of \SI{375.00+-1.06}{\nm}. Systematic errors are shown as shaded areas.}
\label{fig_attenuation_length}
\end{minipage}
\hfill
\begin{minipage}[t]{0.55\linewidth}
\centering
\includegraphics[width=0.99\linewidth]{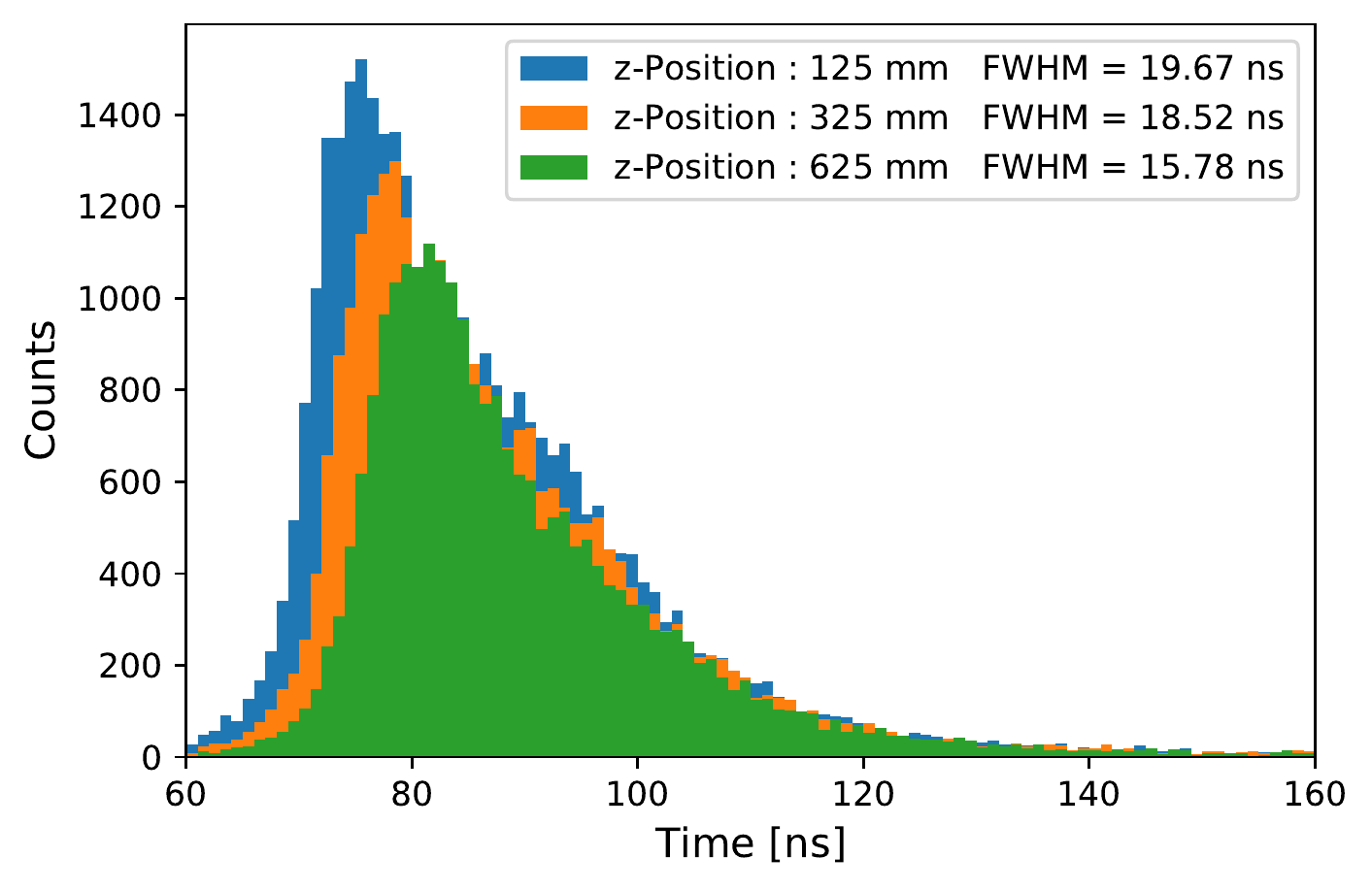}
\caption{The measured transit time spread of a WOM tube with one attached PMT is shown for different $z$-positions along the tube.}
\label{fig_Timing_WOM}
\end{minipage}
\end{figure}

In the absence of light reflected from the other end of the tube, the efficiency measured with the test stand, described in \autoref{subsection_experimental_setup}, can be be identified with the one-sided efficiency in \autoref{eqn:flattened_model}.
The efficiency of the WOM as a function of the $z$-position is determined using the \emph{lock-in} setup (\autoref{eq:eps_meas}) and the \emph{single photon} setup (\autoref{eq:single_eps}) independently for the same WLS-tube.
The results of both measurements overlap within their respective uncertainties. 
In \autoref{fig_attenuation_length} the measured efficiency is shown as a function of the cylinder height $z$ obtained for the 
{\SI{700}{\mm} long and \SI{60}{\mm}} outer diameter prototype quartz tube described in \autoref{sec:intro}.
The tube was coated on the inside at a speed of \SI{~25}{\mm\per\s}.
Using the obtained efficiency as a function of the distance $z$, the \emph{flattened model} (see \autoref{sec:flattened_model}) is used as fit function in conjunction with a normalization constant $N$ as additional fit parameter. The constant $N$ accounts for light losses independent of the position along the tube $z$.

The respective fits lead to attenuation lengths of {\SI{2851 \pm 55}{ \mm} and \SI{3233 \pm 37}{\mm}}
and normalization constants of \SI{89.0 \pm 0.4}{\percent} and \SI{81.0\pm 0.2}{\percent}.
The mean and standard deviation obtained for each PMT measurement result in  an attenuation length of $\lambda_{\rm att}=(3041 \pm 191 )\,\unit{mm}$ and a normalization constant of $N = (84 \pm 4 )\unit{\%}$.
Several factors contribute to the deviation of  $N$ from unity:
\begin{itemize}
    \item Overlap between absorption and emission spectrum of the WLS paint leads to re-absorption of emitted photons with an estimated \unit[0.39]{\%} relative efficiency loss.
    \item Some efficiency is lost due to the inside coating of the tube (see \autoref{eq:emissionPoint} and \autoref{fig:epsilon_TIR}). For the prototype, a reduction in capture efficiency $\epsilon_{\rm TIR}$ compared to outside coating of \unit[4.2]{\%} is obtained. 
    \item Interface losses are calculated using the difference in refractive indices between tube ($n_{\rm Tube} = 1.46$) and glass of the PMT ($n_{\rm PMT} = 1.51$) together with the angular distribution on the end faces from simulation. Averaging between s- and p-polarized transmission yields a relative loss of $1-\epsilon_{\rm IF} = \SI{5.5}{\%}$.
\end{itemize}
Concluding, besides the aspects discussed here, only minor additional efficiency losses may occur and in air $\epsilon_{\rm one-sided}^{\rm meas}=(46.4\pm 4.4)\,\%$ of photons can reach the end of the prototype tube.


\subsection{Transit time spread of photons}
\label{sec:time_resolution}

Using the \emph{single photon} setup, described in \autoref{subsection_experimental_setup}, the transit time spread of photons coupling into the WLS tube is measured as a function of the distance to the PMT $z$. 
Here the arrival time of each individual photon is calculated as the first bin in which the recorded waveform amplitude reaches half of the peak amplitude in the photon event 
{(\emph{constant fraction discriminator})}.  {The transit time spread is the distribution of the individual arrival times.} Measurements are taken on a {\SI{900}{\mm}} long outside coated PMMA tube in increasing distance between PMT and light entry point. 
The  {light} pulser {\cite{Rongen_2018}} has a full width at half maximum (FWHM) of $\sigma =$ \SI{ 0.66}{\ns}
{and the sampling time step of the ADC digitizing the PMT output is \SI{1}{\ns}.} The time resolution of the WOM is a convolution of three effects \cite{Schlickmann:2021}:

\begin{itemize}
    \item time resolution of the PMT which is {measured to be} a Gaussian profile with  $\sigma \sim \SI{6}{\ns}$. 
    \item absorption and re-emission of the WLS paint {measured to be} an exponential decay with $\tau \approx$ \SI{1.5}{\ns} {(in accordance with Ref.} \cite{wls_overview}). {For the lower UV $<280\,{\rm nm}$, where the light needs to be absorbed and shifted by p-Terphenyl first (as mentioned in \autoref{sec:paint}), an additional exponential decay with a time constant of around \SI{1}{\ns} is expected.}
    \item photon trajectory path length distribution inside the tube depending on the absorption length of the material
\end{itemize}

{The final result together with the FWHM of the} distribution {is} shown in \autoref{fig_Timing_WOM}.
The transit time spread of the full module has approximately a FWHM of \SI{\sim 18}{\ns}\footnote{A fit with a symmetric Gaussian gives $\sigma = $\SI{13.5}{\ns}.}. The dominant contribution is caused by the photon propagation in the tube.
A trend towards larger time delays for increasing distances $z$ between PMT and and light entry point for the baseline response time can be observed.


\subsection{Noise\label{sec:noise}}

The background noise of the WOM comprises of the dark noise of the PMT as well as the scintillation noise caused by radioactive decays in the tube glass and optionally the glass of the housing. A climate chamber was used to measure the PMTs dark noise rate at different temperatures. The PMTs were wrapped in black cloth and enclosed in a metal box to prevent impact from background light and radio frequency interference (RFI). To estimate the noise rate of the PMTs, a single photo-electron (SPE) spectrum was obtained from a measurement over time. An exponential together with a Gaussian was fitted to the data and the number of photons under the Gaussian distribution was taken. The noise rate is then estimated by dividing the number of photons by the dead time corrected measurement length in seconds.  
The chosen PMTs have dark noise rates at $-10^\circ\mathrm{C}$ of $r_{\rm DNR}^{-10^\circ}=\left(172.5 \pm 27.7\right)\,\mathrm{Hz}$.

For the choice of the pressure vessel, quartz glass samples from different companies were tested.
For the quartz glass of the prototype \cite{hsq300} a dark noise rate of \SI{6}{\hertz\per\kilo\gram} was measured {at room temperature}, thus $r_{\rm Tube}=4\,\mathrm{Hz}$ for the $700\,\mathrm{g}$ tube and $r_{\rm Housing}=50\,\mathrm{Hz}$ for the $8.3\,\mathrm{kg}$ housing in the prototype design.


\newpage 
\subsection{Effective area\label{sec:aeff}}

{The mean projected effective area is defined here by 

\begin{equation}
    \left< A_{\rm eff}^{\rm proj} \left( \theta, \phi \right)\right> = \frac{N_{\rm det}}{N_{\rm sim}^{\rm plane}}
    \label{eqn:aeff}
\end{equation}

where $N_{\rm sim}^{\rm plane}$ photons are emitted from a plane that is uniformly rotated around the detection device and of these $N_{\rm det}$ will be recorded by the device.}

The effective area of a single PMT with diameter 3.5 inch, see Ref. \cite{hamamatsu}, is obtained by deriving the sensitive projected area for isotropic illumination using the MC simulation described in \autoref{sec:transmission}.
Using a quantum efficiency of 18\% averaged over the WLS emission wavelengths ($400\,\mathrm{nm}$ to $500\,\mathrm{nm}$) and cathode area, the effective area is approximately $ A^{\rm PMT}_{\rm eff}= 12.0\,\mathrm{cm}^2$ in water, ice or air as environment.

For a WOM in the prototype geometry (see \autoref{sec:intro}), all losses from optical propagation into and in the WOM tube, summarized in \autoref{eqn:master_formula}, are {are multiplied with the projected sensitive area. The WLS efficiency is taken from the slide measurement in \autoref{fig_absorption_emission_wls}.
The fraction of captured photons $\varepsilon_{\rm TIR}$ is derived using \autoref{eqn:TIR}.
The attenuation during propagation in the tube $\epsilon_{\mathrm{TP}}$ in dependence of the distance to the tube end is taken from data shown in \autoref{fig_attenuation_length} where the TIR effect is factored out.
The PMT efficiency is chosen as above and the loss at the tube to glue or gel to PMT interfaces is chosen to be $\epsilon_{\mathrm{IF}}=0.95$, compare \autoref{sec:optical_couping}.
}
Then the effective area of the WOM  is approximately $A^{\rm ice}_{\rm eff} = 78.6\,\mathrm{cm}^2$ in water or ice and $A^{\rm air}_{\rm eff} = 126.4 \,\mathrm{cm}^2$ in air  
averaged over the wavelength range of $250\,\mathrm{nm}$ to $600\,\mathrm{nm}$.

\begin{figure}[htb]
\begin{minipage}[t]{0.5\textwidth}
\centering
\includegraphics[width=0.98\textwidth]{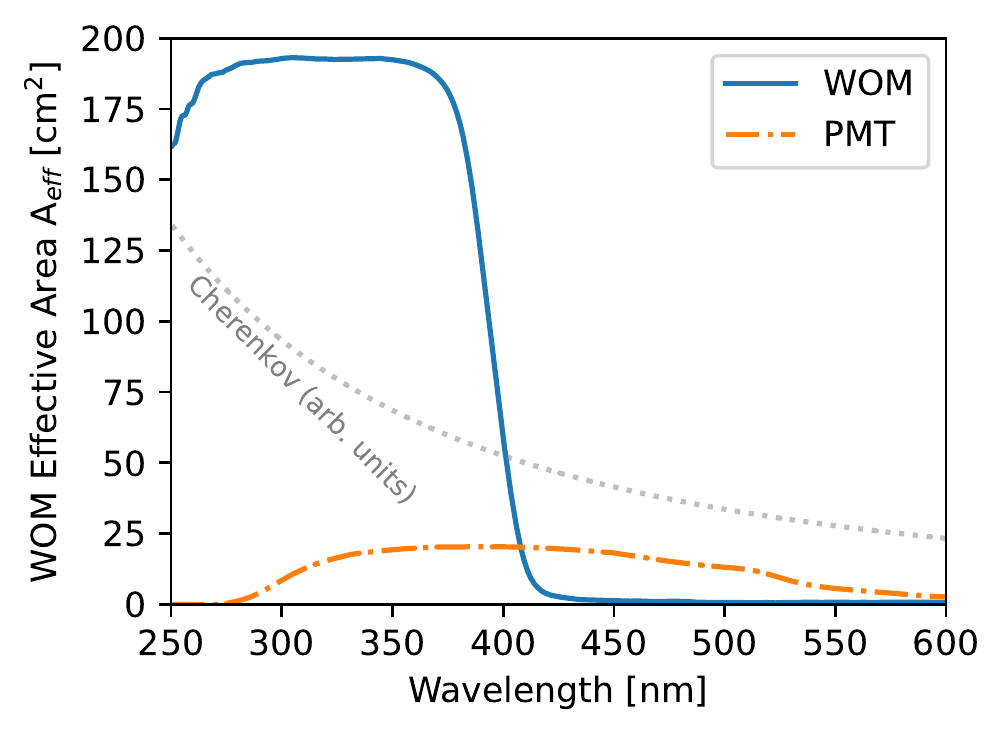}
\caption{Effective areas of the WOM in the prototype design and a single PMT of this ensemble in dependence of wavelength estimated with the MC simulation described in Section \ref{sec:transmission} assuming both devices are deployed in ice. } 
\label{fig:Aeff}
\end{minipage}
\hfill
\begin{minipage}[t]{0.5\textwidth}
\centering
\includegraphics[width=1.0\textwidth]{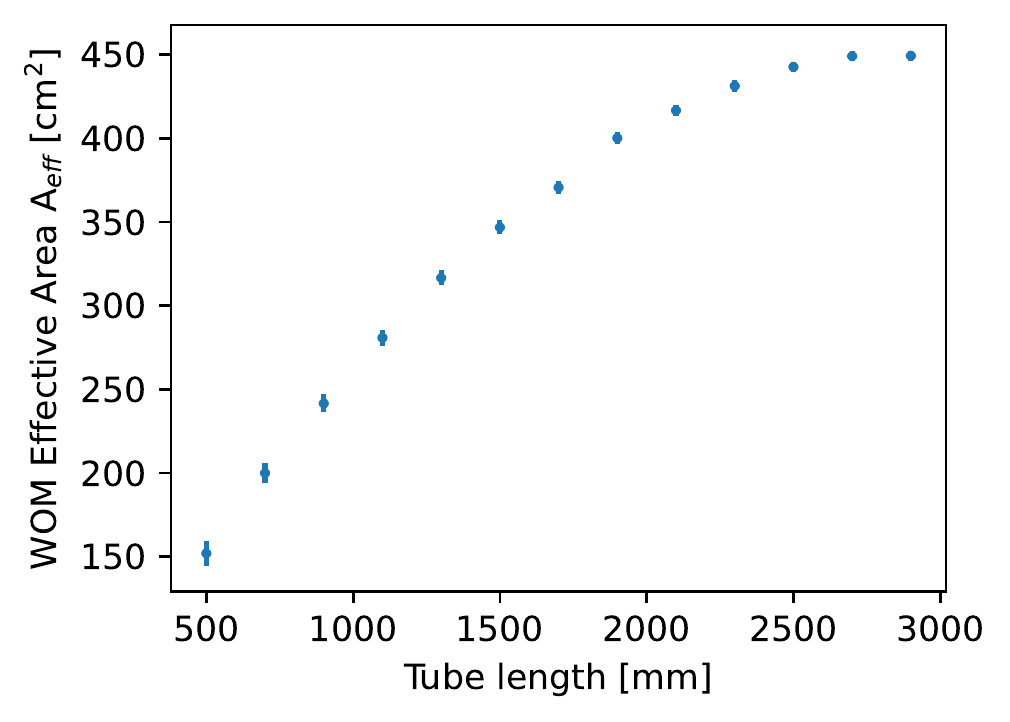}
\caption{Scale of the effective area {at low wavelengths} in ice in dependence of the WOM tube length obtained by MC simulation. Prototype specifications were used for all other properties of the WOM.} 
\label{fig:scale_aefff_len}
\end{minipage}
\end{figure}

In the evaluated design, the effective area of the WOM exceeds the effective area of the single PMT by a factor of approximately $R_{\rm ice} = 6.5$ in ice and $R_{\rm air} = 11.3$ in air.
The wavelength dependent effective area of WOM and a single PMT is shown in \autoref{fig:Aeff}. The Cherenkov spectrum (without attenuation in medium) is shown in parallel, to illustrate the advantage of an enhanced sensitivity at lower wavelength. The ratio of average effective area improves to a factor of approximately $R^{\rm Cher}_{\rm ice}=10.4$ in ice and $R^{\rm Cher}_{\rm air}=18.9$ when weighted with the Cherenkov spectrum. 

The effective area scales linearly with the diameter of the tube. The scaling of the effective area in dependence of the WOM length is {saturating due to the attenuation of photons during propagation in the tube,} shown in \autoref{fig:scale_aefff_len}. 
{In reality the dimensions are further restricted by several limitations: the dip coating station would need to be scaled; glass tubes of appropriate quality are limited in diameter due to manufacturability; high weight and bulkiness of the device complicates handling in the laboratory and during deployment.}

\subsection{Signal-to-noise ratio}

In order to calculate the signal-to-noise ratio (SNR) of the WOM in comparison to a single PMT, the averaged effective area from \autoref{sec:aeff} is used as signal strength and the sum of the noise measurement in \autoref{sec:noise} of all components is used for the denominator. 
In practice the noise of the WOM components would not sum up, at least not for bright signals, because one would add a coincidence requirement for photons hitting both PMTs which should reduce the noise significantly. Here noise rates are summed up conservatively.

For the WOM SNR, this calculation yields improvement factors of approximately $F_{\rm ice}=3.2$ or $F_{\rm air}=5.5$ compared to a single PMT embedded in ice or air. Weighting the effective area with a Cherenkov spectrum (without photon attenuation in medium) would improve the SNR even further. The corresponding improvement factors are approximately $F^{\rm Cher}_{\rm ice}=5.2$ in ice and $F^{\rm Cher}_{\rm air}=8.9$ in air.


\section{Conclusion, applications, and outlook}

In this work, a novel photosensor concept is described in which PMTs are complemented with a tube with wavelength-shifting coating in order to enhance the signal-to-noise ratio. It is demonstrated that consistent coatings can be applied in a simple procedure to glass tubes which allows to effectively convert the UV fraction of the incident spectrum. 
Theoretical studies and experimental work were conducted to understand all features of this concept as well as its performance in detail. In particular, it is shown that in air $\epsilon_{\rm one-sided}^{\rm meas}=\left(46.4 \pm 4.4 \right)\%$  of the converted photons can be detected at the end of the prototype tube. This simple concept can thus be applied to significantly enhance the light collection area of any light sensors. As the WLS tube does not add significantly to the dark noise, for the prototype design this leads to improvement factors in the Cherenkov weighted signal-to-noise ratio of $F_{\rm air}^{\rm Cher}=5.2$ in ice, $F_{\rm ice}^{\rm Cher}=8.9$ in air, respectively. 
As a side effect the sensitivity of the WOMs is improved in the UV which is particularly beneficial for the detection of Cherenkov or scintillation light by matching the emission spectrum of the WLS paint with the PMT wavelength sensitivity. Besides the PMT, the cost of the module is dominated by the coated glass tube while expenses of the paint and glue are insignificant due to the small amounts required for each module.
The transit time spread of detected photons is wider than for bare PMTs because it is smeared predominantly by the propagation time in the tube. 
While the photosensors at the tube ends can be read out independently, only a single photon is generated in the wavelength-shifting process so that in single-photon detection mode no information about the incidence position along the tube can be obtained.

The prototype design does not reach the maximal sensitivity possible in an optimized geometry, due to practical considerations. A wider and longer tube{, if manageable,} would improve the SNR.
Silicone photomultipliers (SiPM) could be chosen for the tube readout in bright conditions, e.g. when deploying in liquid scintillator as detector medium. 
Light guides were evaluated to couple smaller PMTs to wider tubes. However this has stringent theoretical limitations and proved complicated in practice.

\begin{figure}[htb]
\centering
\includegraphics[width=\textwidth]{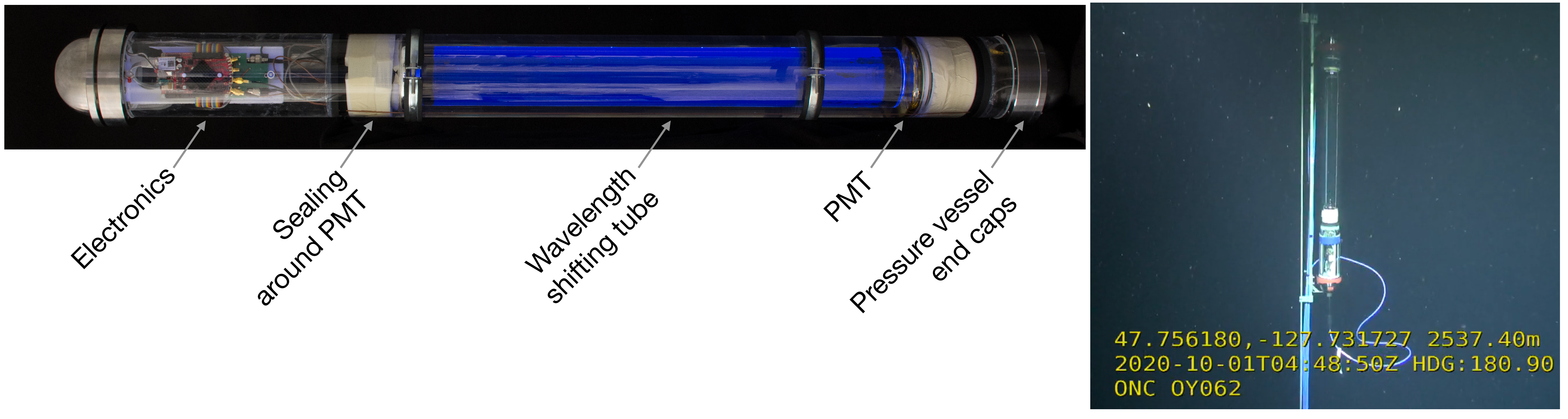}
\caption{{A demonstrator device made in the prototype configuration and photographed under UV illumination (left). The electronics comprises of PMT pulse readout, environmental sensors, and communication. After deployment a photograph was taken by a robot in the Ocean Network Canada. (right). Right picture: Courtesy of ONC.}}
\label{fig:ocn}
\end{figure}

One prototype module enclosed with a vessel was built using simple \emph{of-the-shelf} electronics as \emph{proof-of-concept} following the prototype design described in this work{, see \autoref{fig:ocn}}. Hydrogel \cite{Bubeck:2020} was used as filling material to improve the optical properties of the module. The module was deployed in the Canadian sea at a depth of $2539\,\textrm{m}$ within the Ocean Network Canada \cite{Rea:2021hjd}. It operated stably at the given conditions and providing data of the local bio-luminescence.

Another application, here prototypes are already at hand, is the envisioned SHiP experiment \cite{Ehlert:2018pke, SHIP_new}.
SiPM are coupled to the WLS tubes which are to be deployed in the 
{veto chamber of the experiment}.

For the upcoming IceCube Upgrade, which extends the low-energy infill of IceCube, a WOM with an improved design, adjusted to the environmental needs is designed. Twelve modules with $1300\,\mathrm{mm}$ long pressure vessels, with 5 inch PMTs, and customized electronics are developed and will be deployed \cite{UpgradeWOM}.


\authorcontributions{\added[id=AP]{Investigation and methodology: Benjamin Bastian-Querner, Lucas Binn, Sebastian Böser, Jannes Brostean-Kaiser, Dustin Hebecker, Klaus Helbing, Timo Karg, Lutz Köpke, Marek Kowalski, Peter Peiffer, Anna Pollmann, John Rack-Helleis, Martin Rongen, Lea Schlickmann, Florian Thomas and Anna Vocke;
Supervision: Sebastian Böser, Klaus Helbing, Timo Karg, Lutz Köpke and Marek Kowalski;
Writing – review and editing: Sebastian Böser, Anna Pollmann and John Rack-Helleis.}}

\funding{We acknowledge the support from the following agencies: Bundesministerium f\"ur Bildung und Forschung (BMBF), Deutsche Forschungsgemeinschaft (DFG) in particular through the clusters of Excellence PRISMA and PRISMA$^{+}$, Initiative and Networking Fund of the Helmholtz Association, \added[id=AP]{Deutsches Elektronen-Synchrotron DESY}. \added[id=AP]{We acknowledge support from the Open Access Publication Fund of the University of Wuppertal.}}

{\added[id=AP]{\textbf{Institutional Review Board Statement:} Not applicable.}}

{\added[id=AP]{\textbf{Informed Consent Statement:} {Not applicable.}}

\acknowledgments{ %
We acknowledge the fruitful discussion with all the individuals involved in establishing the WOM as sensor for the SHiP surround background tagger (SBT).

We are very grateful to ONC and the P-ONE Collaboration for the deployment of the WOM prototype as part of the STRAW-b installation

We would also like to thank Andreas Best from the 
Max-Planck-Institute for polymer research for the profilometer measurements which allowed us to estimate the thickness of our WLS coatings. 

We are thankful to Jochen Schreiner and Hardy Simgen of the Max Planck-Institute for Nuclear physics in Heidelberg, as well as Christoph Düllmann from the Institute of Nuclear Physics in Mainz for the radioactive impurity measurements on our glass samples.

We further acknowledge the numerous studies conducted by students in the context of the WOM which are not explicitly presented in this publication. In particular, these are: 
Kristina Sand$^{2}$,
Peter Falke$^{1}$,
Carl Fösig$^{2}$,
Esther del Pino Rosendo$^{2}$,
Jan Weldert$^{2}$, 
Elisa Lohfink$^{2}$,
Christian Matthe$^{2}$,
Lucas Binn$^{2}$,
Sandra Gerlach$^{2}$,
Alexandre Portier$^{4}$,
\added[id=AP]{Yuriy Popovych$^{2}$},
Daniel Popper$^{2}$,
Ronja Schnur$^{2}$,
Andreas Looft$^{3}$,
Jakob Beise$^{3}$,
Maximillian Bubeck$^{2}$,
Maximillian Thiel$^{2}$,
\added[id=AP]{Nich Jannis Schmeisser$^{5}$}%
; from the institutions:
$^{1}$Universtität Bonn, 
$^{2}$University of Mainz, 
$^{3}$DESY \deleted[id=AP]{Zeuthen}, 
$^{4}$PHELMA Grenoble, 
\added[id=AP]{$^{5}$University of Wuppertal}.
%
}

\conflictsofinterest{The authors declare no conflict of interest.}

\externalbibliography{yes}
\reftitle{References}
\bibliography{thebibliography.bib}

\begin{thebibliography}{-------}
\providecommand{\natexlab}[1]{#1}

\bibitem[Aartsen and et~al(2017)]{icecube}
Aartsen, M.G.; et~al., {\bf IceCube} Collaboration.
\newblock {The IceCube Neutrino Observatory: Instrumentation and Online
  Systems}.
\newblock {\em JINST} {\bf 2017}, {\em 12},~P03012,
  \href{http://xxx.lanl.gov/abs/1612.05093}{{\normalfont
  [arXiv:astro-ph.IM/1612.05093]}}.

\bibitem[Fukuda and et~al(2003)]{FUKUDA2003418}
Fukuda, S.; et~al., {\bf Super-Kamiokande} Collaboration.
\newblock The Super-Kamiokande detector.
\newblock {\em Nucl. Instrum. Meth. A} {\bf 2003}, {\em 501},~418--462.
\newblock
  doi:{\changeurlcolor{black}\href{https://doi.org/10.1016/S0168-9002(03)00425-X}{\detokenize{10.1016/S0168-9002(03)00425-X}}}.

\bibitem[Alimonti and et~al(2009)]{ALIMONTI2009568}
Alimonti, G.; et~al., {\bf Borexino} Collaboration.
\newblock The Borexino detector at the Laboratori Nazionali del Gran Sasso.
\newblock {\em Nucl. Instrum. Meth. A} {\bf 2009}, {\em 600},~568--593.
\newblock
  doi:{\changeurlcolor{black}\href{https://doi.org/10.1016/j.nima.2008.11.076}{\detokenize{10.1016/j.nima.2008.11.076}}}.

\bibitem[Boger and et~al(2000)]{SNO:1999crp}
Boger, J.; et~al., {\bf SNO} Collaboration.
\newblock {The Sudbury neutrino observatory}.
\newblock {\em Nucl. Instrum. Meth. A} {\bf 2000}, {\em 449},~172--207,
  \href{http://xxx.lanl.gov/abs/9910016}{{\normalfont
  [arXiv:nucl-ex/9910016]}}.

\bibitem[An and et~al(2016)]{Juno}
An, F.; et~al., {\bf JUNO} Collaboration.
\newblock {Neutrino physics with JUNO}.
\newblock {\em J. Phys. G: Nucl. Part. Phys.} {\bf 2016}, {\em 43},~030401.
\newblock
  doi:{\changeurlcolor{black}\href{https://doi.org/10.1088/0954-3899/43/3/030401}{\detokenize{10.1088/0954-3899/43/3/030401}}}.

\bibitem[Aprile and et~al(2012)]{APRILE2012573}
Aprile, E.; et~al., {\bf XENON} Collaboration.
\newblock The XENON100 dark matter experiment.
\newblock {\em Astropart. Phys.} {\bf 2012}, {\em 35},~573--590.
\newblock
  doi:{\changeurlcolor{black}\href{https://doi.org/10.1016/j.astropartphys.2012.01.003}{\detokenize{10.1016/j.astropartphys.2012.01.003}}}.

\bibitem[Akerib and et~al(2013)]{AKERIB2013111}
Akerib, D.; et~al., {\bf LUX} Collaboration.
\newblock The Large Underground Xenon (LUX) experiment.
\newblock {\em Nucl. Instrum. Meth. A} {\bf 2013}, {\em 704},~111--126.
\newblock
  doi:{\changeurlcolor{black}\href{https://doi.org/10.1016/j.nima.2012.11.135}{\detokenize{10.1016/j.nima.2012.11.135}}}.

\bibitem[Aalbers and et~al(2016)]{Aalbers_2016}
Aalbers, J.; et~al., {\bf DARWIN} Collaboration.
\newblock {DARWIN}: towards the ultimate dark matter detector.
\newblock {\em J. Cosmol. Astropart. Phys.} {\bf 2016}, {\em 2016},~017--017.
\newblock
  doi:{\changeurlcolor{black}\href{https://doi.org/10.1088/1475-7516/2016/11/017}{\detokenize{10.1088/1475-7516/2016/11/017}}}.

\bibitem[Tseung and Tolich(2011)]{LAB}
Tseung, H.W.C.; Tolich, N.
\newblock {Ellipsometric measurements of the refractive indices of linear
  alkylbenzene and EJ-301 scintillators from 210 to 1000 nm}.
\newblock {\em Phys. Scr.} {\bf 2011}, {\em 84},~035701,
  \href{http://xxx.lanl.gov/abs/1105.2101}{{\normalfont
  [arXiv:physics.optics/1105.2101]}}.

\bibitem[Her(2021)]{hsq300}
Heraeus.
\newblock {\em HSQ 300},  2021.
\newblock Datasheet online available:
  \url{https://www.heraeus.com/media/media/hca/doc_hca/products_and_solutions_8/solids/Solids_HSQ300_330MF_EN.pdf}.

\bibitem[Ham(2021)]{hamamatsu}
Hamamatsu.
\newblock {\em R14689},  2021.
\newblock Datasheet online available:
  \url{https://www.hamamatsu.com/eu/en/product/type/R14689/index.html}.

\bibitem[Elj(2021)]{eljen}
Eljen Technology.
\newblock {\em EJ-550},  2021.
\newblock Datasheet online available:
  \url{https://eljentechnology.com/products/accessories/ej-550-ej-552}.

\bibitem[Thomas(2019)]{Thomas:2019}
Thomas, F.
\newblock {Light propagation simulation for the Wavelength-shifting Optical
  Module on CUDA GPUs}.
\newblock Master's thesis, {Johannes Gutenberg University Mainz}, Germany,
  2019.
\newblock
  \url{https://butler.physik.uni-mainz.de/icecube/thesis/master_Florian_Thomas.pdf}.

\bibitem[Her(2021)]{transmission}
Heraeus.
\newblock {\em Transmission calculator for optical applications},  2021.
\newblock URL incl. settings:
  \url{https://www.heraeus.com/en/hca/fused_silica_quartz_knowledge_base_1/t_calc_1/transmission_calc_opt/transmission_calculator_opt.html?chartIndex=2&selection=suprasil_311_312\%2Csuprasil_1_2a\%2Csuprasil_2b\%2Cspectrosil_2000&thickness=10&rangeX=120\%2C4500}.

\bibitem[Qua(2022{\natexlab{a}})]{QuantumDesign}
Quantum Design.
\newblock {\em Arc light sources 50 - 150 W arc light source},  2022.
\newblock Datasheet online available:
  \url{https://qd-europe.com/fileadmin/Mediapool/products/lightsources/en/LQ_50_150_w_arc_light_source_en.pdf}.

\bibitem[Qua(2022{\natexlab{b}})]{MSH300}
Quantum Design.
\newblock {\em Monochromators Monochromator MSH-300 with variable slit},  2022.
\newblock Datasheet online available:
  \url{https://qd-europe.com/fileadmin/Mediapool/products/Bentham/_pdf/MSH_300_with_variable_slit.pdf}.

\bibitem[Zur(2021)]{zurich}
Zurich Instruments.
\newblock {\em zi MFLI Lock in Amplifier},  2021.
\newblock Datasheet online available:
  \url{https://www.zhinst.com/sites/default/files/documents/2021-12/zi_mfli_leaflet_v2.pdf}.

\bibitem[Ham(2022)]{Photodiode}
Hamamatsu Photonics.
\newblock {\em Si photodiodes with BNC connector},  2022.
\newblock Datasheet online available:
  \url{https://www.hamamatsu.com/resources/pdf/ssd/s2281_series_kspd1044e.pdf}.

\bibitem[Tho(2021)]{Thorlabs}
Thorlabs.
\newblock {\em Thorlabs liquid Light guide},  2021.
\newblock Datasheet online available:
  \url{https://www.thorlabs.com/drawings/fe93bd8d318e7307-7BA1AD90-E649-0882-7162D968CDA653D9/LLG5-4T-SpecSheet.pdf}.

\bibitem[Rack-Helleis(2019)]{Rack-Helleis:2019}
Rack-Helleis, J.
\newblock Efficiency determination of the Wavelength-shifting Optical Module
  (WOM).
\newblock Master's thesis, {Johannes Gutenberg University Mainz}, Germany,
  2019.
\newblock
  \url{https://butler.physik.uni-mainz.de/icecube/thesis/master_John_Rack-Helleis.pdf}.

\bibitem[Rongen and Schaufel(2018)]{Rongen_2018}
Rongen, M.; Schaufel, M.
\newblock Design and evaluation of a versatile picosecond light pulser.
\newblock {\em JINST} {\bf 2018}, {\em 13},~P06002.
\newblock
  doi:{\changeurlcolor{black}\href{https://doi.org/10.1088/1748-0221/13/06/p06002}{\detokenize{10.1088/1748-0221/13/06/p06002}}}.

\bibitem[Tel(2021)]{teledyne}
Teledyne.
\newblock {\em SP Devices ADQ 14},  2021.
\newblock Datasheet online available:
  \url{https://www.spdevices.com/documents/datasheets/19-adq14-datasheet/file}.

\bibitem[Kuzniak and Szelc(2020)]{wls_overview}
Kuzniak, M.; Szelc, A.M.
\newblock Wavelength Shifters for Applications in Liquid Argon Detectors.
\newblock {\em Instruments} {\bf 2020}, pp. 4 --5.
\newblock
  doi:{\changeurlcolor{black}\href{https://doi.org/10.3390/instruments5010004}{\detokenize{10.3390/instruments5010004}}}.

\bibitem[Hebecker(2014)]{Hebecker:2014}
Hebecker, D.
\newblock Developement of a single photon detector with wavelength shifting and
  light guiding technology.
\newblock Master's thesis, {University of Bonn}, Germany,  2014.
\newblock
  \url{https://www-zeuthen.desy.de/~hebecked/Publications_etc./Master_Thesis/Dustin_hebecker_master_thesis.pdf}.

\bibitem[Pre(2021)]{paraloid}
Preservation Equipment.
\newblock {\em Paraloid B72},  2021.
\newblock Datasheet online available:
  \url{https://www.preservationequipment.com/files//4ba8f3dc-85c1-44e4-9237-a3db00db1ef4/Paraloid%20B72%20Use.pdf}.

\bibitem[Beise(2019)]{Beise:2019}
Beise, J.
\newblock Transport losses in light guides for the WOM application.
\newblock Bachelor's thesis, {Humboldt-University Berlin}, Germany,  2019.

\bibitem[Car(2021)]{mucasol}
Carl Roth GmbH + Co. KG.
\newblock {\em Mucasol Universalreiniger},  2021.
\newblock Datasheet online available:
  \url{https://www.carlroth.com/de/de/reinigungsmittel-fuer-ultraschallgeraete/universalreiniger-mucasol/p/1a3l.1}.

\bibitem[Abdel-Mottaleb and Ahmed(2009)]{Abdel-Mottaleb:2009vo}
Abdel-Mottaleb, M.S.; Ahmed, R.M.
\newblock Optical Study on Polymethyl methacrylate/Polyvinyl acetate Blends.
\newblock {\em Int. J. Photoenergy} {\bf 2009}, {\em 2009},~150389.
\newblock
  doi:{\changeurlcolor{black}\href{https://doi.org/10.1155/2009/150389}{\detokenize{10.1155/2009/150389}}}.

\bibitem[Brinker(2013)]{Brinker2013}
Brinker, C.J., Dip Coating.
\newblock In {\em Chemical Solution Deposition of Functional Oxide Thin Films};
  Springer,  2013; pp. 233--261.
\newblock
  doi:{\changeurlcolor{black}\href{https://doi.org/10.1007/978-3-211-99311-8_10}{\detokenize{10.1007/978-3-211-99311-8_10}}}.

\bibitem[Rio and Boulogne(2017)]{RIO2017100}
Rio, E.; Boulogne, F.
\newblock Withdrawing a solid from a bath: How much liquid is coated?
\newblock {\em Adv. Colloid Interface Sci.} {\bf 2017}, {\em 247},~100--114.
\newblock
  doi:{\changeurlcolor{black}\href{https://doi.org/10.1016/j.cis.2017.01.006}{\detokenize{10.1016/j.cis.2017.01.006}}}.

\bibitem[{Derjaguin}(1993)]{Derjaguin:43d}
{Derjaguin}, B.
\newblock {On the thickness of the liquid film adhering to the walls of a
  vessel after emptying}.
\newblock {\em Prog. Surf. Sci} {\bf 1993}, {\em 43},~134--137.
\newblock
  doi:{\changeurlcolor{black}\href{https://doi.org/10.1016/0079-6816(93)90022-N}{\detokenize{10.1016/0079-6816(93)90022-N}}}.

\bibitem[Binn(2018)]{Binn:2018}
Binn, L.S.
\newblock Charakterisierung von dünnen wellenlängenschiebenden Schichten.
\newblock Bachelor's thesis, {University of Mainz}, Germany,  2018.
\newblock
  \url{https://butler.physik.uni-mainz.de/icecube/thesis/bachelor_Lucas_Binn.pdf}.

\bibitem[Nor(2021)]{norland2}
Norland.
\newblock {\em NOA 148H},  2021.
\newblock Datasheet online available:
  \url{https://www.norlandprod.com/adhesives/NOA148.html}.

\bibitem[Falke(2014)]{Falke:2014}
Falke, P.
\newblock Entwicklung eines Lichtkonzentrators basierend auf einer Hohlzylinder
  Geometrie.
\newblock Bachelor's thesis, {Universität Bonn}, Germany,  2014.

\bibitem[Schnur(2020)]{Schnur:2020}
Schnur, R.
\newblock Optimierung des adiabatischen Lichtleiters für das
  Wavelength-shifting Optical Module.
\newblock Bachelor's thesis, {Johannes Gutenberg-Universität Mainz}, Germany,
  2020.
\newblock
  \url{https://butler.physik.uni-mainz.de/icecube/thesis/bachelor_Ronja_Schnur.pdf}.

\bibitem[Hebecker(2021)]{Hebecker:2021}
Hebecker, D.
\newblock Development of a single photon detector using wavelength-shifting and
  light-guiding technology.
\newblock PhD thesis, {Humbold-Universität Berlin}, Germany,  2021.
\newblock
  \url{https://edoc.hu-berlin.de/handle/18452/23885?locale-attribute=de}.

\bibitem[Schlickmann(2021)]{Schlickmann:2021}
Schlickmann, L.
\newblock {Zeitantwort des Wellenl\"angenschiebenden Optischen Moduls (WOM)}.
\newblock Master's thesis, {Johannes Gutenberg University Mainz}, Germany,
  2021.
\newblock
  \url{https://butler.physik.uni-mainz.de/icecube/thesis/bachelor_SchlickmannLea.pdf}
  (in German).

\bibitem[Bubeck(2020)]{Bubeck:2020}
Bubeck, M.
\newblock Developement of a Wavelength-shifting Optical Module.
\newblock Master's thesis, {Johannes Gutenberg University Mainz}, Germany,
  2020.
\newblock
  \url{https://butler.physik.uni-mainz.de/icecube/thesis/master_Maximilian_Bubeck.pdf},
  page 19.

\bibitem[Rea and et~al(2021)]{Rea:2021hjd}
Rea, I.C.; et~al.
\newblock {P-ONE second pathfinder mission: STRAW-b}.
\newblock {\em PoS} {\bf 2021}, {\em ICRC2021},~1092.
\newblock
  doi:{\changeurlcolor{black}\href{https://doi.org/10.22323/1.395.1092}{\detokenize{10.22323/1.395.1092}}}.

\bibitem[Ehlert \em{et~al.}(2019)Ehlert, Hollnagel, Korol, Korzenev, Lacker,
  Mermod, Schliwinski, Shihora, Venkova, and Wurm]{Ehlert:2018pke}
Ehlert, M.; Hollnagel, A.; Korol, I.; Korzenev, A.; Lacker, H.; Mermod, P.;
  Schliwinski, J.; Shihora, L.; Venkova, P.; Wurm, M.
\newblock {Proof-of-principle measurements with a liquid-scintillator detector
  using wavelength-shifting optical modules}.
\newblock {\em JINST} {\bf 2019}, {\em 14},~P03021.
\newblock
  doi:{\changeurlcolor{black}\href{https://doi.org/10.1088/1748-0221/14/03/P03021}{\detokenize{10.1088/1748-0221/14/03/P03021}}}.

\bibitem[Ahdida and et~al(2021)]{SHIP_new}
Ahdida, C.; et~al.
\newblock {The SHiP experiment at the proposed CERN SPS Beam Dump Facility},
  2021,  \href{http://xxx.lanl.gov/abs/2112.01487}{{\normalfont
  [arXiv:physics.ins-det/2112.01487]}}.

\bibitem[Rack-Helleis \em{et~al.}(2021)Rack-Helleis, Pollmann, and
  Rongen]{UpgradeWOM}
Rack-Helleis, J.; Pollmann, A.; Rongen, M.
\newblock {The Wavelength-shifting Optical Module (WOM) for the IceCube
  Upgrade}.
\newblock {\em PoS} {\bf 2021}, {\em ICRC2021},~1038,
  \href{http://xxx.lanl.gov/abs/2107.10194}{{\normalfont
  [arXiv:astro-ph.HE/2107.10194]}}.

\end{thebibliography}

\end{document}